\documentclass[letterpaper,twocolumn,10pt]{article}
\usepackage{styles/usenix}

\newcommand{\usenix}[1]{\ignorespaces}
\newcommand{\arxiv}[1]{#1}

\newcommand{\papertitle}[0]{\Large\textbf{Expected Exploitability: Predicting the Development of Functional Vulnerability Exploits}}
\newcommand{\papertitlearxiv}[0]{\Large\textbf{Technical Report - Expected Exploitability: Predicting the Development of Functional Vulnerability Exploits}}

\usepackage{combelow}
\usepackage{amsmath}
\usepackage{amssymb}
\usepackage[normalem]{ulem}   
\usepackage{soul}       
\usepackage{xurl}                  
\usepackage{calc}                 
\usepackage{shadow}             
\usepackage{fancyvrb}           
\usepackage{xspace}             
\usepackage{ifthen}               
\usepackage{comment}            
\usepackage{multirow}           
\usepackage{dcolumn}            
\usepackage{colortbl}
\usepackage{tabularx}           
\usepackage{mdwtab}             
\usepackage{mdwlist}      
\usepackage{ifpdf}                
\usepackage{rotating}           
\usepackage{listings}
\usepackage{wrapfig}            
\usepackage{pdfpages}     
\usepackage{graphicx}           
\usepackage{subcaption}
\usepackage{hyperref}
\usepackage{parcolumns}
\usepackage{enumitem}

\usepackage{titling}
\date{}

\usepackage{titlesec}

\usepackage[table, subgroups]{svn-multi} 

\svnidlong
{$HeadURL: https://mc2.umiacs.umd.edu/svn/SDSatUMD/papers/2019_functional_exploits/functional_exploits.tex $}
{$LastChangedDate: 2019-09-14 06:14:20 -0400 (Sat, 14 Sep 2019) $}
{$LastChangedRevision: 7503 $}
{$LastChangedBy: osuciu $}

\def\paperversionmajor{1}          
\def\paperversionminor{\svnrev}    

\def\monthName#1{\ifcase#1\or
  January\or February\or March\or April\or May\or June\or
  July\or August\or September\or October\or November\or December\fi}

\usenix{\hypersetup{pdftitle={\papertitle}}}
\arxiv{\hypersetup{pdftitle={\papertitlearxiv}}}
\hypersetup{pdfproducer={v. \paperversionmajor.\paperversionminor}}
\ifpdf
\fi

\newcommand{\TODO}[2][]{\noindent\colorbox{lime}{
    \color{red}
    \parbox{\minof{.97\hsize}{\widthof{#1} * \real{1.15} + \widthof{ #2}}}{\textbf{#1} #2}
    \par}
}

\sethlcolor{lime}
\newcommand{\todo}[2][]{
    \textcolor{red}{\hl{\textbf{#1} #2}}
}

\renewcommand{\TODO}[2][]{}
\renewcommand{\todo}[2][]{}

\newcounter{hypothesis}                                     

\newcommand{\topic}[1]{\vspace{2pt}\noindent\textbf{#1.}}

\newcommand{\ee}{\texttt{EE}}

\graphicspath{ {figures/} }
\usepackage[functional]{styles/usenixbadges}

\arxiv{

}
\usenix{
\pagestyle{empty}
}
\microtypecontext{spacing=nonfrench}
\begin{document}



\usenix{\title{\papertitle}}
\arxiv{\title{\papertitlearxiv}}


\author{%
  Octavian Suciu,
  Connor Nelson\textsuperscript{\textdagger}, 
  Zhuoer Lyu\textsuperscript{\textdagger}, 
  Tiffany Bao\textsuperscript{\textdagger},
  Tudor Dumitra\cb{s}
  \\
  \textit{University of Maryland, College Park}\\
  \textsuperscript{\textdagger}\textit{Arizona State University}
}

\maketitle
\thispagestyle{empty}
%
%

\begin{abstract}
\noindent 

Assessing the exploitability of software vulnerabilities at the time of disclosure is difficult and error-prone, as features extracted via technical analysis by existing metrics are poor predictors for exploit development.
Moreover, exploitability assessments suffer from a class bias because ``not exploitable'' labels could be inaccurate.

To overcome these challenges, we propose a new metric, called Expected Exploitability (\texttt{EE}), which reflects, over time, the likelihood that functional exploits will be developed. 
Key to our solution is a time-varying view of exploitability, a departure from existing metrics. 
This allows us to learn \texttt{EE} using data-driven techniques from artifacts published after disclosure, such as technical write-ups and proof-of-concept exploits, for which we design novel feature sets.

This view also allows us to investigate the effect of the label biases on the classifiers.
We characterize the noise-generating process for exploit prediction, showing that our problem is subject to the most challenging type of label noise, and propose techniques to learn \texttt{EE} in the presence of noise.

On a dataset of 103,137 vulnerabilities, we show that \texttt{EE} increases precision from 49\% to 86\% over existing metrics, including two state-of-the-art exploit classifiers, while its precision substantially improves over time.
We also highlight the practical utility of \texttt{EE} for predicting imminent exploits and prioritizing critical vulnerabilities. 

We develop \texttt{EE} into an online platform which is publicly available at \href{https://exploitability.app/}{https://exploitability.app/}.
\end{abstract}

\usenix{
\pagestyle{empty}
}


%
%

\section{Introduction}
\label{sec:intro}

Weaponized exploits have a disproportionate impact on security, as highlighted in 2017 by the WannaCry~\cite{talosintelligence_wannacry} and NotPetya~\cite{crowdstrike_notpetya} worms that infected millions of computers worldwide.
Their notorious success was in part due to the use of weaponized exploits. 
The cyber-insurance industry regards such contagious malware, which propagates automatically by exploiting software vulnerabilities, as the leading risk for incurring large losses from cyber attacks~\cite{rms_risk}.
%
%
At the same time, the rising bar for developing weaponized exploits~\cite{szekeres2013sok} pushed black-hat developers to focus on exploiting only 5\% of the known vulnerabilities~\cite{WEIS-19-Exploits}.
To prioritize mitigation efforts in the industry, to make optimal decisions in the government's Vulnerabilities Equities Process~\cite{VEPdocument}, and to gain a deeper understanding of the research opportunities to prevent exploitation, we must evaluate each vulnerability's ease of exploitation.

Despite significant advances in defenses~\cite{szekeres2013sok}, exploitability assessments remain elusive because we do not know which vulnerability \textit{features} predict exploit development. 
%
For example, 
expert recommendations for prioritizing patches~\cite{redmonmag:wannacry,qualys:wannacry} initially omitted CVE-2017-0144, the vulnerability later exploited by WannaCry and NotPetya.
While one can prove exploitability by developing an exploit, it is challenging to establish non-exploitability, as this requires reasoning about state machines with an unknown state space and emergent instruction semantics~\cite{dullien2017weird}. 
This results in a \textit{class bias} of exploitability assessments, as we cannot be certain that a ``not exploitable'' label is accurate.

We address these two challenges through a metric called \textit{Expected Exploitability} (\texttt{EE}).
Instead of deterministically labeling a vulnerability as ``exploitable'' or ``not exploitable'', our metric continuously estimates \textit{over time} the likelihood that a \textit{functional exploit} will be developed, based on historical patterns for similar vulnerabilities.
%
Functional exploits go beyond proof-of-concepts (POCs) to achieve the full security impact prescribed by the vulnerability.
While functional exploits are readily available for real-world attacks, 
we aim to predict their development and not their use in the wild, which depends on many other factors besides exploitability~\cite{USENIX-Security-2015,xiao2018patching,WEIS-19-Exploits}.
%

Key to our solution is a time-varying view of exploitability, a departure from the existing vulnerability scoring systems such as CVSS~\cite{cvss3guide}, which are not designed to take into account new information (e.g., new exploitation techniques, leaks of weaponized exploits) that becomes available after the scores are initially computed~\cite{eiram2013exploitability}.
By systematically comparing a range of prior and novel features, we observe that artifacts published after vulnerability disclosure can be good predictors for the development of exploits, but their timeliness and predictive utility varies.
This highlights limitations of prior features and a qualitative distinction between predicting functional exploits and related tasks.
%
For example, prior work uses the existence of public PoCs as an exploit predictor~\cite{Tavabi2018DarkEmbedEP,jacobs2019exploit,WEIS-19-Exploits}.
%
However, PoCs are designed to trigger the vulnerability by crashing or hanging the target application and often are not directly weaponizable; we observe that this leads to many false positives for predicting functional exploits.
%
In contrast, we discover that certain PoC characteristics, such as the code complexity, are good predictors, because triggering a vulnerability is a necessary step for every exploit, making these features causally connected to the difficulty of creating functional exploits.
We design techniques to extract features at scale, from PoC code written in 11 programming languages, which complement and improve upon the precision of previously proposed feature categories.
We then learn \texttt{EE} from the useful features using data-driven methods, which have been successful in predicting other incidents, e.g., vulnerabilities that are exploited in the wild~\cite{USENIX-Security-2015,xiao2018patching,WEIS-19-Exploits}, data breaches~\cite{liu2015cloudy} or website compromises~\cite{DBLP:conf/uss/SoskaC14}. 

However, learning to predict exploitability could be derailed by a biased ground truth.
%
%
Although prior work had acknowledged this challenge for over a decade~\cite{DBLP:conf/kdd/BozorgiSSV10,USENIX-Security-2015},
no attempts were made to address it.
This problem, known in the machine-learning literature as \textit{label noise}, can significantly degrade the performance of a classifier. 
The time-varying view of exploitability allows us to uncover the root causes of label noise: exploits could be published only after the data collection period ended, which in practice translates to wrong negative labels.
This insight allows us to characterize the noise-generating process for exploit prediction and propose a technique to mitigate the impact of noise when learning \texttt{EE}.


%
%
%
%
%
In our experiments on 103,137 vulnerabilities, \texttt{EE} significantly outperforms static exploitability metrics and prior state-of-the art exploit predictors, increasing the precision from 49\% to 86\% one month after disclosure. 
Using our label noise mitigation technique, the classifier performance is minimally affected even if evidence about 20\% of exploits is missing.
Furthermore, by introducing a metric to capture vulnerability prioritization efforts, we show that \texttt{EE} requires only 10 days from disclosure to approach its peak performance.
We show \texttt{EE} has practical utility, by providing timely predictions for imminent exploits, even when public PoCs are unavailable.
Moreover, when employed on scoring 15 critical vulnerabilities, \texttt{EE}
places them above 96\% of non-critical ones, compared to only 49\% for existing metrics.


%
%

In summary, our contributions are as follows:
\begin{itemize}[noitemsep, topsep=0pt]
    \item We propose a time-varying view of exploitability based on which we design Expected Exploitability (\texttt{EE}), a metric to learn and continuously estimate the likelihood of functional exploits over time.
    \item We characterize the noise-generating process systematically affecting exploit prediction, and propose a domain-specific technique to learn \texttt{EE} in the presence of label noise.
    \item We explore the timeliness and predictive utility of various artifacts, proposing new and complementary features from PoCs, and developing scalable feature extractors for them.
    \item We perform \arxiv{three} \usenix{two} case studies to investigate the practical utility of \texttt{EE}, showing that it can qualitatively improve prioritization strategies based on exploitability.
\end{itemize}

\texttt{EE} is available as an online platform at \href{https://exploitability.app/}{https://exploitability.app/} and described in the Artifact Appendix~\ref{artifact_appendix:0}.

\section{Problem Overview}
\label{sec:problem}

We define exploitability as the likelihood that \emph{a functional exploit}, which fully achieves the mandated security impact, will be developed for a vulnerability.
Exploitability reflects the technical difficulty of exploit development, and it does not capture the feasibility of lunching exploits against targets in the wild~\cite{USENIX-Security-2015,xiao2018patching,WEIS-19-Exploits}, which is influenced by additional factors (e.g., patching delays, network defenses, attacker choices).


While an exploit represents conclusive proof that a vulnerability is exploitable if it can be generated, proving non-exploitability is significantly more challenging~\cite{dullien2017weird}.
%
%
%
Instead, mitigation efforts are often guided by vulnerability scoring systems, which aim to capture exploitation difficulty, such as:

\begin{enumerate}[noitemsep, topsep=0pt]
\item \textit{NVD CVSS}~\cite{cvss3guide}, a mature scoring system with its Exploitability metrics intended to reflect the ease and technical means by which the vulnerability can be exploited. 
The score encodes various vulnerability characteristics, such as the required access control, complexity of the attack vector and privilege levels, into a numeric value between 0 and 4 (0 and 10 for CVSSv2), with 4 reflecting the highest exploitability.

\item \textit{Microsoft Exploitability Index}~\cite{MS:ExploitabilityIndex}, a vendor-specific score assigned by experts using one of four values to communicate to Microsoft customers the likelihood of a vulnerability being exploited~\cite{eiram2013exploitability}.
%

\item \textit{RedHat Severity}~\cite{RedHat:SeverityRating}, similarly encoding the difficulty of exploiting the vulnerability by complementing CVSS with expert assessments based on vulnerability characteristics specific to the RedHat products.
\end{enumerate}

The estimates provided by these metrics are often inaccurate, as highlighted by prior work~\cite{Allodi12:VulnerabilityScores,allodi2014comparing,riskbasedsecurity:CVSS1,Reuters:ExploitabilityIndexCritique,eiram2013exploitability} and by our analysis in Section~\ref{sec:empirical_results}.
For example, CVE-2018-8174, an exploitable Internet Explorer vulnerability, received a CVSS exploitability score of 1.6, placing it below 91\% of vulnerability scores.
Similarly, CVE-2018-8440, an exploited vulnerability affecting Windows 7 through 10 was assigned score of 1.8.


\begin{figure}[t]
\centering
\includegraphics[width=0.48\textwidth]{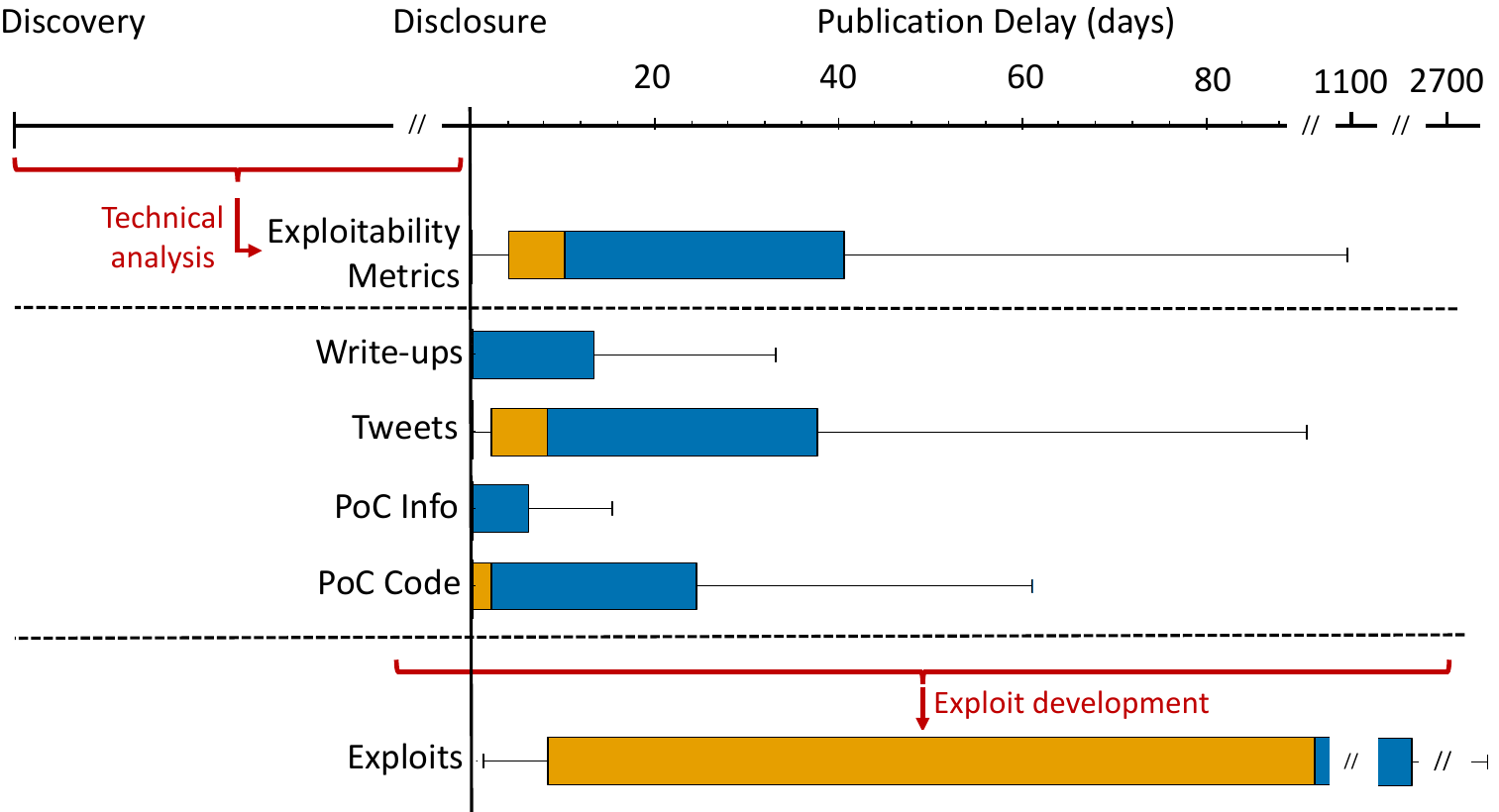}
\caption{Vulnerability timeline highlighting publication delay for different artifacts and the CVSS Exploitability metric. The box plot delimits the $25^{th}$, $50^{th}$ and $75^{th}$ percentiles, and the whiskers mark $1.5$ times the interquartile range.
}
\label{fig:timeline}
\end{figure}

To understand why these metrics are poor at reflecting exploitability, we highlight the typical timeline of a vulnerability in Figure~\ref{fig:timeline}.
The exploitability metrics depend on a technical analysis which is performed before the vulnerability is disclosed publicly, and which \textit{considers the vulnerability statically and in isolation}.

However, we observe that public disclosure is followed by the publication of various vulnerability artifacts such as write-ups and PoCs containing code and additional technical information about the vulnerability, and social media discussions around them.
These artifacts often provide meaningful information about the likelihood of exploits.
For CVE-2018-8174 it was reported that the publication of technical write-ups was a direct cause for exploit development in exploit kits~\cite{bleepingcomp-ekits}, while a PoC for CVE-2018-8440 has been determined to trigger exploitation in the wild within two days~\cite{zdnet-win10disclosureexploit}.
The examples highlight that \textit{existing metrics fail to take into account useful exploit information available only after disclosure and they do not update over time}. 

Figure~\ref{fig:timeline} plots the publication delay distribution for different artifacts released after disclosure, according to our data analysis described in Section~\ref{sec:empirical_results}.
Data shows not only that these artifacts become available soon after disclosure,\footnote{Interestingly, some social media posts appear before disclosure. These "leaks from coordinated disclosure" were investigated by prior work~\cite{USENIX-Security-2015}.} providing opportunities for timely assessments, but also that \textit{static exploitability metrics}, such as CVSS, \textit{are often not available at the time of disclosure}.

\topic{Expected Exploitability}
\label{sec:expectedexp}
The problems mentioned above suggest that the evolution of exploitability over time can be described by a stochastic process.
At a given point in time, exploitability is a random variable $E$ encoding the probability of observing an exploit.
$E$ assigns a probability $0.0$ to the subset of vulnerabilities that are provably unexploitable, and $1.0$ to vulnerabilities with known exploits.
Nevertheless, the true distribution $\mathsf{E}$ generating $E$ is not available at scale, and instead we need to rely on a noisy version $\mathsf{E}^{train}$, as we will discuss in Section~\ref{sec:challenges}.
This implies that in practice $E$ has to be approximated from the available data, by computing the likelihood of exploits, which estimates the expected value of exploitability. 
We call this measure \textit{Expected Exploitability} (\texttt{EE}).
\texttt{EE} can be learned from historical data using supervised machine learning and can be used to assess the likelihood of exploits for new vulnerabilities \textit{before functional exploits are developed or discovered}.

\section{Challenges}
\label{sec:challenges}
We recognize three challenges in utilizing supervised techniques for learning, evaluating and using \texttt{EE}.

\topic{Extracting features from PoCs} 
Prior work investigated the existence of PoCs as predictors for exploits, repeatedly showing that they lead to a poor precision~\cite{Allodi12:VulnerabilityScores,jacobs2019exploit,Tavabi2018DarkEmbedEP}.
However, PoCs are designed to trigger the vulnerability, a step also required in a functional exploit.
As a result, the structure and complexity of the PoC code can reflect exploitation difficulty directly: a complex PoC implies that the functional exploit will also be complex.
To fully leverage the predictive power of PoCs, we need to capture these characteristics.
We note that while public PoCs have a lower coverage compared to other artifact types, they are broadly available privately because they are often mandated when vulnerabilities are reported~\cite{hackerone:disclosure}.

Extracting features using NLP techniques from prior exploit prediction work~\cite{DBLP:conf/kdd/BozorgiSSV10,USENIX-Security-2015,WEIS-19-Exploits} is not sufficient, because code semantics differs from that of natural language.
Moreover, PoCs are written in different programming languages and are often malformed programs~\cite{Allodi12:VulnerabilityScores,mu2018understanding}, combining code with free-form text, which limits the applicability of existing program analysis techniques.
PoC feature extraction therefore requires text and code separation, and robust techniques to obtain useful code representations.

\topic{Understanding and mitigating label noise}
    Prior work found that the labels available for training have biases~\cite{DBLP:conf/kdd/BozorgiSSV10,USENIX-Security-2015}, but to our knowledge no prior attempts were made to link this issue to the problem of label noise.
    The literature distinguishes two models of non-random label noise, according to the generating distribution: class-dependent and feature-dependent~\cite{frenay2013classification}. 
    The former assumes a uniform label flipping probability among all instances of a class, while the latter assumes that noise probability also depends on individual features of instances. 
    If $\mathsf{E}^{train}$ is affected by label noise, the test time performance of the classifier could suffer.  

    By viewing exploitability as time-varying, it becomes immediately clear that exploit evidence datasets are prone to class-dependent noise.
    This is because exploits might not yet be developed or be kept secret.
    Therefore, a subset of vulnerabilities believed not to be exploited are in fact wrongly labeled at any given point in time. 
    
    In addition, prior work noticed that individual vendors providing exploit evidence have uneven coverage of the vulnerability space (e.g., an exploit dataset from Symantec would not contain Linux exploits because the platform is not covered by the vendor)~\cite{USENIX-Security-2015}, suggesting that noise probability might be dependent on certain features.
    The problem of feature-dependent noise is much less studied~\cite{patrini2017making}, and discovering the characteristics of such noise on real-world applications is considered an open problem in machine learning~\cite{frenay2013classification}.
    
    Exploit prediction therefore requires an empirical understanding of both the type and effects of label noise, as well as the design of learning techniques to address it.
    
\topic{Evaluating the impact of time-varying exploitability}
    While some post-disclosure artifacts are likely to improve classification, publication delay might affect their utility as timely predictions.
    Our \texttt{EE} evaluation therefore needs to use metrics which highlight potential trade-offs between timeliness and performance.
    Moreover, the evaluation needs to test whether our classifier can capitalize on artifacts with high predictive power available before functional exploits are discovered, and whether \texttt{EE} can capture the imminence of certain exploits.
    Finally, we need to demonstrate the practical utility of \texttt{EE} over existing static metrics,
    in real-world scenarios involving vulnerability prioritization. 


\topic{Goals and Non-Goals}
\label{sec:goals-non-goals}
Our goal is to estimate \texttt{EE} for a broad range of vulnerabilities, by addressing the challenges listed above. 
Moreover, we aim to provide estimates that are both accurate and robust: they should predict the development of functional exploits \textit{better} than the existing scoring systems and despite \textit{inaccuracies in the ground truth}. 
%
The closest work to our goal is DarkEmbed~\cite{Tavabi2018DarkEmbedEP}, which uses natural language models trained on underground forum discussions to predict the availability of exploits. 
In contrast, we aim to predict functional exploits from public information, a more difficult task as we lack direct evidence of black-hat exploit development. 

We do not aim to generate functional exploits automatically. 
We also do not aim to analyze the code of existing exploits in order to assess how close they are to becoming functional. 
Instead, we aim to quantify the exploitability of known vulnerabilities objectively, by predicting whether functional exploits will be developed for them.
While we aim to make our exploit predictor robust to systematic label noise, we do not attempt to improve the model's robustness to adversarial examples~\cite{USENIX-Security-2015}, as techniques for achieving this have been widely studied elsewhere~\cite{chakraborty2018adversarial}.
Finally, we do not aim to predict which vulnerabilities are likely to be exploited in the wild, in real-world attacks, because this likelihood is influenced by additional factors (e.g., attacker choices, patching delays) that are out of scope for this paper.

%
%

\section{Data Collection}
\label{sec:data_collection}

In this section we describe the methods used to collect the data used in our paper, as well as the techniques for discovering various timestamps in the lifecycle of vulnerabilities.



\subsection{Gathering Technical Information}  
\label{sec:data_gathering}

We use the CVEIDs to identify vulnerabilities, because it is one of the most prevalent and cross-referenced public vulnerability identification systems.
Our collection contains vulnerabilities published between January 1999 and March 2020.

\topic{Public Vulnerability Information}
We begin by collecting information about the vulnerabilities targeted by the PoCs from the National Vulnerability Database (NVD)~\cite{nvd}.
NVD adds vulnerability information gathered by analysts, including textual descriptions of the issue, product and vulnerability type information, as well as the CVSS score.
Nevertheless, NVD only contains high-level descriptions. 
%
In order to build a more complete coverage of the technical information available for each vulnerability, we search for external references in several public sources.
We use the Bugtraq~\cite{securityfocus} and IBM X-Force Exchange~\cite{xforce}, vulnerability databases which provide additional textual descriptions for the vulnerabilities.
We also use Vulners~\cite{vulners}, a database that collects in real time textual information from vendor advisories, security bulletins, third-party bug trackers and security databases. 
We filter out the reports that mention more than one CVEID, as it would be challenging to determine which particular one is discussed.
In total, our collection contains 278,297 documents from 76 sources, referencing 102,936 vulnerabilities.
We refer to these documents as \textit{write-ups}, which, together with the NVD textual information and vulnerability details, provide a broader picture of the \textit{technical information} publicly available for vulnerabilities.

\topic{Proof of Concepts (PoCs)}
We collect a dataset of public PoCs by scraping ExploitDB~\cite{exploitdb}, Bugtraq~\cite{securityfocus} and Vulners~\cite{vulners}, three popular vulnerability databases that contain exploits aggregated from multiple sources.
Because there is substantial overlap across these sources, but the formatting of the PoCs might differ slightly, we remove duplicates using a content hash that is invariant to such minor whitespace differences.
We preserve only the 48,709 PoCs that are linked to CVEIDs, which correspond to 21,849 distinct vulnerabilities.


\topic{Social Media Discussions} 
We also collect social media discussions about vulnerabilities from Twitter, by gathering tweets mentioning CVE-IDs between January 2014 and December 2019.
We collected 1.4 million tweets for 52,551 vulnerabilities by continuously monitored the Twitter Filtered Stream API~\cite{Twitter:api}, using the approach from our prior work~\cite{USENIX-Security-2015}.
While the Twitter API does not sample returned tweets, short offline periods for our platform caused some posts to be lost. 
By a conservative estimate using the lost tweets which were later retweeted, our collection contains over 98\% of all public tweets about these vulnerabilities.

\topic{Exploitation Evidence Ground Truth} Because we are not aware of any comprehensive dataset of evidence about developed exploits, we aggregate evidence from multiple public sources.

We begin from the Temporal CVSS score, which tracks the status of exploits and the confidence in these reports. 
The Exploit Code Maturity component has four possible values: "Unproven", "Proof-of-Concept", "Functional" and "High".
The first two values indicate that the exploit is not practical or not functional, while the last two values indicate the existence of autonomous or functional exploits that work in most situations. 
Because the Temporal score is not updated in NVD, we collect it from two reputable sources: IBM X-Force Exchange~\cite{xforce} threat sharing platform and the Tenable Nessus~\cite{Tenable:Nessus} vulnerability scanner.
Both scores are used as inputs to proprietary severity assessment solutions: the former is used by IBM in one of their cloud offerings~\cite{IBM:RiskRating}, while the latter is used by Tenable as input to commercial vulnerability prioritization solutions~\cite{Tenable:RiskRating}.
We use the labels "Functional" and "High" as evidence of exploitation, as defined by the official CVSS Specification~\cite{cvss3guide}, obtaining 28,009 exploited vulnerabilities.
We further collect evidence of 2,547 exploited vulnerabilities available in three commercial exploitation tools: Metasploit~\cite{metasploit}, Canvas~\cite{d2canvas} and D2~\cite{d2pack}. 
We also scrape the Bugtraq~\cite{securityfocus} exploit pages, and create NLP rules to extract evidence for 1,569 functional exploits.
Examples of indicative phrases are: \textit{"A commercial exploit is available."}, \textit{"A functional exploit was demonstrated by researchers."}.


We also collect exploitation evidence that results from exploitation in the wild. 
Starting from a dataset collected from Symantec in our prior work~\cite{USENIX-Security-2015}, we update it by scraping Symantec's Attack Signatures~\cite{attack_signatures} and Threat Explorer~\cite{sym_threatexplorer}. 
We then aggregate labels extracted using NLP rules (matching e.g., \textit{"... was seen in the wild."}) from scrapes of Bugtraq~\cite{securityfocus}, Tenable~\cite{tenable}, Skybox~\cite{skybox} and AlienVault OTX~\cite{otx}.
In addition, we use the Contagio dump~\cite{contagio} which contains a curated list of exploits used by exploit kits.
These sources were reported by prior work as reliable for information about exploits in the wild~\cite{Nappa15:VulnerabilityPatching,USENIX-Security-2015,WEIS-19-Exploits}. 
Overall, 4,084 vulnerabilities are marked as exploited in the wild.

While exact development time for most exploits is not available, we drop evidence if we cannot confirm they were published within one year after vulnerability disclosure, simulating a historical setting.
Our ground truth, consisting of 32,093 vulnerabilities known to have functional exploits, therefore reflects a lower bound for the number of exploits available, which translates to class-dependent label noise in classification, issue that we evaluate in Section~\ref{sec:results}.

\subsection{Estimating Lifecycle Timestamps}  
\label{sec:lifecycle_estimation}

Vulnerabilities are often published in NVD at a later date than their public disclosure~\cite{Bilge12:ZeroDay, li2017large}.
%
%
We estimate the public disclosure dates for the vulnerabilities in our dataset by selecting the minimum date among all write-ups in our collection and the publication date in NVD, in line with prior research~\cite{shahzad:12,li2017large}.
This represents the earliest date when expected exploitability can be evaluated. 
We validate our estimates for the disclosure dates by comparing them to two independent prior estimates~\cite{shahzad:12,li2017large}, on the 67\% of vulnerabilities which are also found in the other datasets. 
We find that the median date difference between the two estimates is 0 days, and our estimates are an average of 8.5 days earlier than prior assessments. 
Similarly, we estimate the time when PoCs are published as the minimum date among all sources that shared them, and we confirm the accuracy of these dates by verifying the commit history in exploit databases that use version control.

To assess whether \texttt{EE} can provide timely warnings, we need the dates for the emergence of functional exploits and attacks in the wild.
Because the sources of exploit evidence do not share the dates when exploits were developed, we estimate these dates from ancillary data.
For the exploit toolkits, we collect the earliest date when exploits are reported in the Metasploit and Canvas platforms. 
For exploits in the wild, we use the dates of first recorded attacks, from prior work~\cite{USENIX-Security-2015}.
Across all exploited vulnerabilities, we also crawl VirusTotal~\cite{virustotal}, a popular threat sharing platform, for the timestamps when exploit files were first submitted.
Finally, we estimate exploit availability as the earliest date among the different sources, excluding vulnerabilities with zero-day exploits.
Overall, we discovered this date for 10\% (3,119) of the exploits.
These estimates could result in label noise, because exploits might sometimes be available earlier, e.g., PoCs that are easy to weaponize. In Section~\ref{sec:casestudy} we measure the impact of such label noise on the \texttt{EE} performance. 



\subsection{Datasets}  
\label{sec:datasets}
We create three datasets that we use throughout the paper to evaluate \texttt{EE}.
\textbf{\texttt{DS1}} contains all 103,137 vulnerabilities in our collection that have at least one artifact published within one year after disclosure. We use this to evaluate the timeliness of various artifacts, compare the performance of \texttt{EE} with existing baselines, and measure the predictive power of different categories of features.
The second dataset, \textbf{\texttt{DS2}}, contains 21,849 vulnerabilities that have artifacts across all different categories within one year. This is used to compare the predictive power of various feature categories, observe their improved utility over time, and to test their robustness to label noise.
The third one, \textbf{\texttt{DS3}} contains 924 out of the 3,119 vulnerabilities for which we estimated the exploit emergence date, and which are disclosed during our classifier deployment period described in Section~\ref{sec:design_exploit_predictor}. These are used to evaluate the ability of \texttt{EE} to distinguish imminent exploit.

\section{Empirical Observations}
\label{sec:empirical_results}

\begin{figure}
     \centering
     \begin{subfigure}[b]{0.15\textwidth}
         \centering
         \includegraphics[height=1.6in]{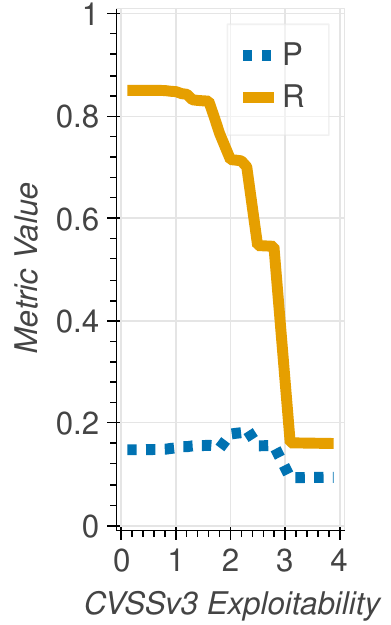}
     \end{subfigure}
     \begin{subfigure}[b]{0.15\textwidth}
         \centering
         \includegraphics[height=1.6in]{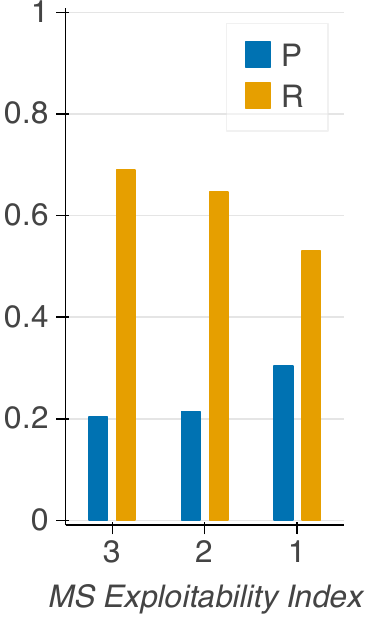}
     \end{subfigure}
     \begin{subfigure}[b]{0.15\textwidth}
         \centering
         \includegraphics[height=1.6in]{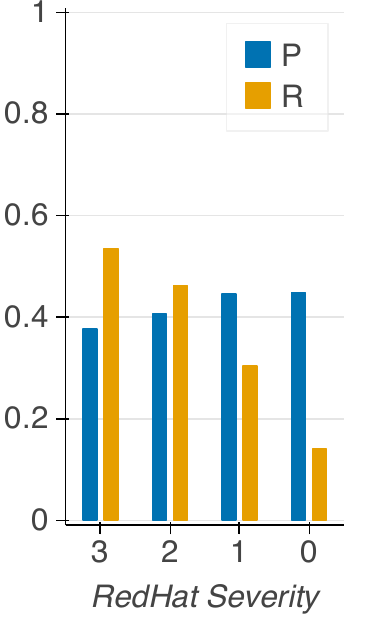}
     \end{subfigure}
        \caption{Performance of existing severity scores at capturing exploitability. We report both precision (P) and recall (R). The numerical score values are ordered by increasing severity.
        }
        \label{fig:baseline}
\end{figure}

\arxiv{
\begin{figure}
     \centering
     \begin{subfigure}[b]{0.47\textwidth}
         \centering
         \includegraphics[height=1.15in]{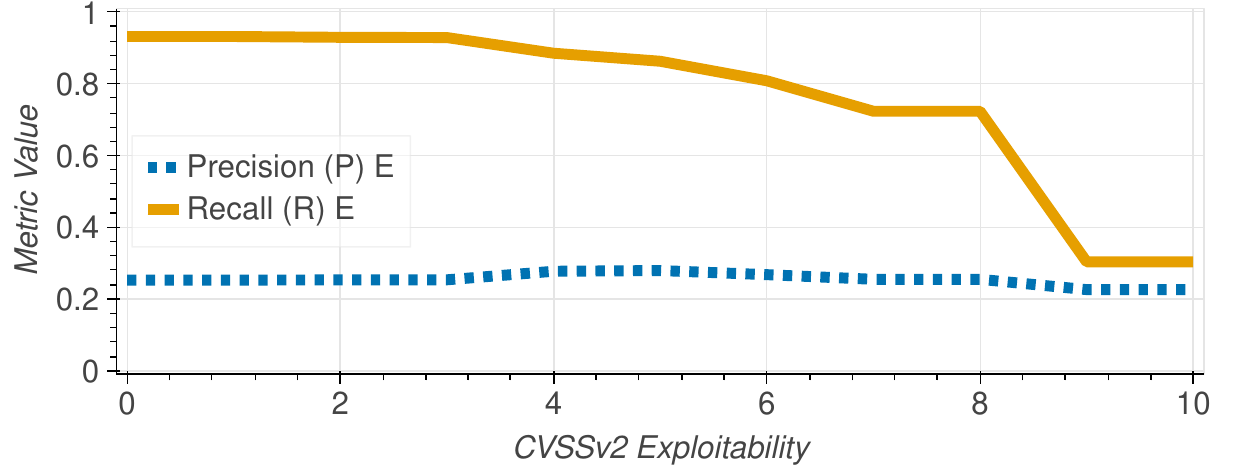}
     \end{subfigure}
        \caption{Performance of CVSSv2 at capturing exploitability. We report both precision (P) and recall (R).}
        \label{fig:baselinev2}
\end{figure}
}

We start our analysis with three empirical observations on \texttt{DS1}, which guide the design of our system for computing \texttt{EE}.

\topic{Existing scores are poor predictors}\label{sec:method}
First, we investigate the effectiveness of three vulnerability scoring systems, described in Section~\ref{sec:problem}, for predicting exploitability. 
Because these scores are widely used, we will utilize them as baselines for our prediction performance; our goal for \texttt{EE} is to improve this performance substantially. 
%
As the three scores do not change over time, we utilize a threshold-based decision rule, which predicts that all vulnerabilities with scores greater or equal than the threshold are exploitable. 
By varying the threshold across the entire score range, and using all the vulnerabilities in our dataset, we evaluate their \textit{precision (P)}: the fraction of predicted vulnerabilities that have functional exploits within one year from disclosure, and \textit{recall (R)}: the fraction of exploited vulnerabilities that are identified within one year.
%

Figure~\ref{fig:baseline} reports these performance metrics.
%
It is possible to obtain $R=1$ by marking all vulnerabilities as exploitable, but this affects $P$ because many predictions would be false positives.
%
For this reason, for all the scores, $R$ decreases as we raise the severity threshold for prediction. 
However, obtaining a high $P$ is more difficult. 
For CVSSv3 Exploitability, $P$ does not exceed 0.19, regardless of the detection threshold, and some vulnerabilities do not have scores assigned to them.
\usenix{
In the technical report~\cite{suciu2021expected} we evaluate CVSSv2, which yields similar results.
}
\arxiv{
CVSSv2 also exhibits a very poor precision, as illustrated in Figure~\ref{fig:baselinev2}.
}

When evaluating the Microsoft Exploitability Index on the 1,100 vulnerabilities for Microsoft products in our dataset disclosed since the score inception in 2008,
we observe that the maximum precision achievable is 0.45. 
The recall is also lower because the score is only computed on a subset of vulnerabilities~\cite{eiram2013exploitability}.

On the 3,030 vulnerabilities affecting RedHat products, we observe a similar trend for the proprietary severity metric, where precision does not exceed 0.45. 

These results suggest that \textit{the three existing scores predict exploitability with $>50$\% false positives}.
This is compounded by the facts that (1) \textit{some scores are not computed for all vulnerabilities, owing to the manual effort required}, which introduces false negative predictions; (2) \textit{the scores do not change, even if new information becomes available}; and (3) \textit{not all the scores are available at the time of disclosure}, meaning that the recall observed operationally soon after disclosure will be lower, as highlighted in the next section.


\begin{figure}[t]
    \begin{subfigure}{.24\textwidth}
        \centering
        \includegraphics[height=1.8in]{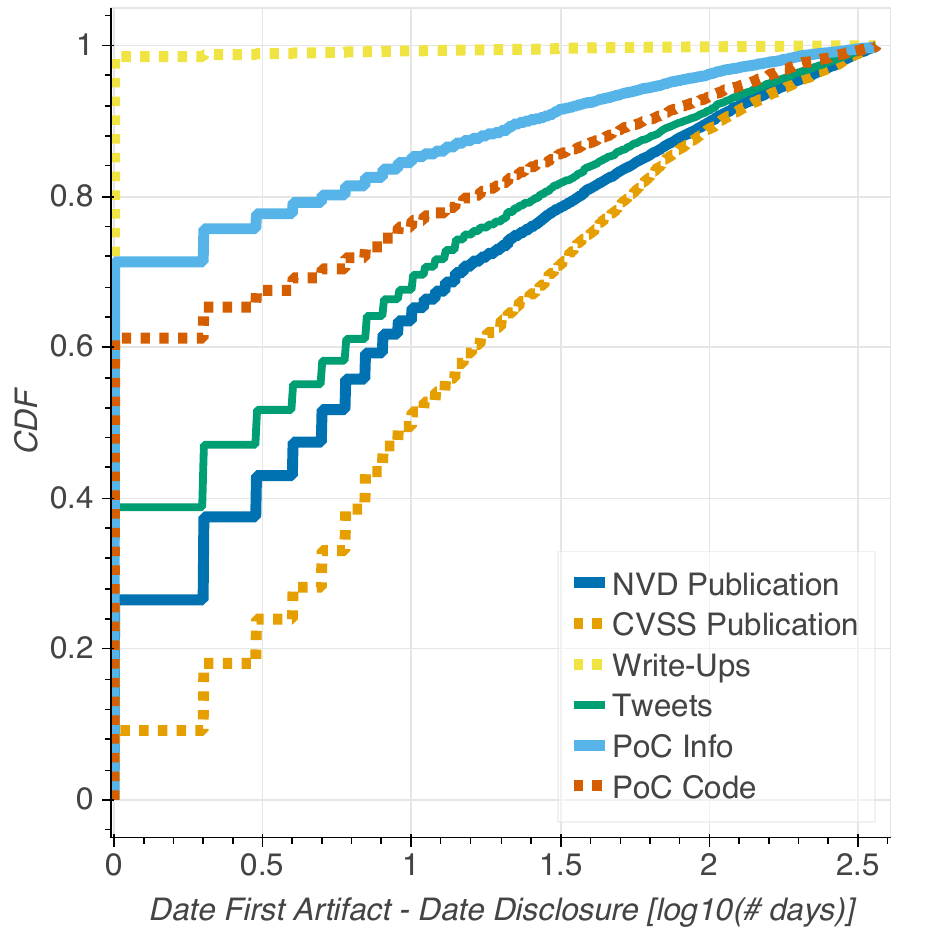}
        \vspace{-0.2in}
        \caption{}
        \label{fig:cdf_disclosure}
    \end{subfigure}%
    \begin{subfigure}{.24\textwidth}
        \centering
        \includegraphics[height=1.8in]{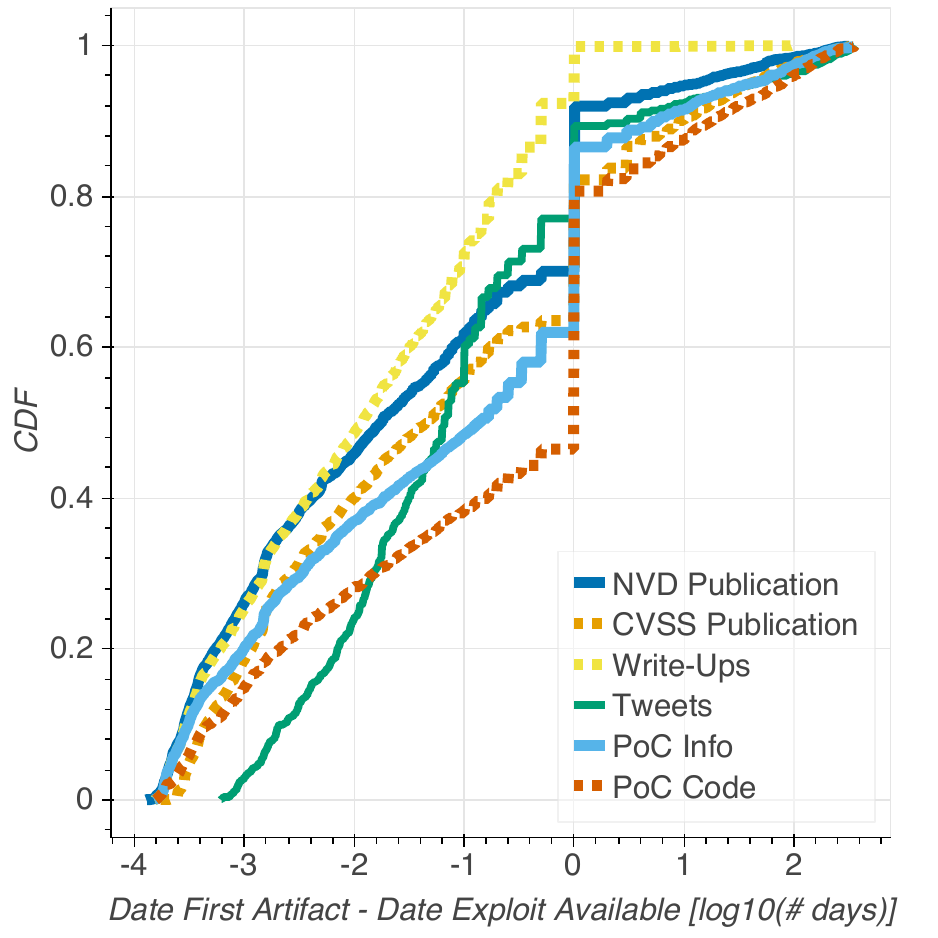}
        \vspace{-0.2in}
        \caption{}
        \label{fig:cdf_earlyprediction}
    \end{subfigure}%
\caption{(a) Number of days after disclosure when vulnerability artifacts are first published. (b) Difference between the availability of exploits and availability of other artifacts. The day differences are in logarithmic scale.}
\label{fig:cdf_disclosure_earlypred}
\end{figure}

\topic{Artifacts provide early prediction opportunities}
To assess the opportunities for early prediction, we look at the publication timing for certain artifacts from the vulnerability lifecycle.
In Figure~\ref{fig:cdf_disclosure_earlypred}(a), we plot, across all vulnerabilities, the earliest point in time after disclosure when the first write-ups are published, they are added to NVD, their CVSS and technical analysis are published in NVD, their first PoCs are released and when they are first mentioned on Twitter.
The publication delay distribution for all collected artifacts is available in Figure~\ref{fig:timeline}.

Write-ups are the most widely available ones at the time of disclosure, suggesting that vendors prefer to disclose vulnerabilities through either advisories or third-party databases.
However, many PoCs are also published early: 71\% of vulnerabilities have a PoC on the day of disclosure.
In contrast, only 26\% of vulnerabilities in our dataset are added to NVD on the day of disclosure, and surprisingly, only 9\% of the CVSS scores are published at disclosure.
This result suggests that \textit{timely exploitability assessments require looking beyond NVD, using additional sources of technical vulnerability information, such as the write-ups and PoCs}.
This observation drives our feature engineering from Section~\ref{sec:feature_engineering}.

Figure~\ref{fig:cdf_disclosure_earlypred}(b) highlights the day difference between the dates when the exploits become available and the availability of the artifacts from public vulnerability disclosure.
For more than 92\% of vulnerabilities, write-ups are available before the exploits become available. 
We also find that the 62\% of PoCs are available before this date, while 64\% of CVSS assessments are added to NVD before. 
Overall, the availability of exploits is highly correlated with the emergence of other artifacts, indicating an \textit{opportunity to infer the existence of functional exploits as soon as, or before, they become available}. 

\begin{table*}[t]
\renewcommand{\arraystretch}{1.2}
\setlength{\tabcolsep}{5pt}
\centering
\resizebox{\textwidth}{!}{
\begin{tabular}{ |c|| c | c | c | c | c | c || c | c | c | c | c | c |}
  \hline
 & \multicolumn{6}{c||}{\small\strut\textbf{Functional Exploits}} & \multicolumn{6}{c|}{\small\strut\textbf{Exploits in the Wild}} \\ \hline
 & \small{Tenable} & \small{X-Force} & \small{Metasploit} & \small{Canvas} & \small{Bugtraq} & \small{D2} & \small{Symantec} & \small{Contagio} & \small{Alienvault} & \small{Bugtraq} & \small{Skybox} & \small{Tenable}\\ \hline
\small{CWE-79} & \tiny{\checkmark} & \tiny{\checkmark} & \tiny{\checkmark} & \tiny{\checkmark} & \tiny{\checkmark} & \tiny{\checkmark} & \tiny{\checkmark} & \tiny{\checkmark} & \tiny{\checkmark} & \tiny{\checkmark} & \tiny{\checkmark} & \tiny{\checkmark (0.006)}
\\ \hline
\small{CWE-94} & \tiny{\checkmark} & \tiny{\checkmark} & \tiny{\checkmark} & \tiny{\checkmark} & \tiny{\checkmark} & \tiny{\checkmark} & \tiny{\checkmark} & \tiny{\checkmark} & \tiny{\checkmark} & \tiny{\checkmark} & \tiny{\checkmark} & \tiny{\text{\sffamily X} (1.000)}
\\ \hline
\small{CWE-89} & \tiny{\checkmark} & \tiny{\checkmark} & \tiny{\checkmark} & \tiny{\checkmark} & \tiny{\checkmark} & \tiny{\text{\sffamily X} (1.000)} & \tiny{\checkmark} & \tiny{\checkmark} & \tiny{\checkmark} & \tiny{\checkmark} & \tiny{\checkmark} & \tiny{\text{\sffamily X} (0.284)}
\\ \hline
\small{CWE-119} & \tiny{\checkmark} & \tiny{\checkmark} & \tiny{\checkmark} & \tiny{\checkmark} & \tiny{\checkmark} & \tiny{\checkmark} & \tiny{\checkmark} & \tiny{\checkmark} & \tiny{\checkmark} & \tiny{\checkmark} & \tiny{\checkmark} & \tiny{\checkmark (0.001)}
\\ \hline
\small{CWE-20} & \tiny{\checkmark} & \tiny{\checkmark} & \tiny{\checkmark} & \tiny{\checkmark} & \tiny{\checkmark} & \tiny{\checkmark} & \tiny{\checkmark} & \tiny{\text{\sffamily X} (1.000)} & \tiny{\checkmark} & \tiny{\checkmark (0.002)} & \tiny{\checkmark} & \tiny{\text{\sffamily X} (1.000)}
\\ \hline
\small{CWE-22} & \tiny{\checkmark} & \tiny{\text{\sffamily X} (0.211)} & \tiny{\checkmark} & \tiny{\text{\sffamily X} (1.000)} & \tiny{\text{\sffamily X} (1.000)} & \tiny{\checkmark} & \tiny{\text{\sffamily X} (1.000)} & \tiny{\text{\sffamily X} (1.000)} & \tiny{\text{\sffamily X} (0.852)} & \tiny{\text{\sffamily X} (1.000)} & \tiny{\text{\sffamily X} (1.000)} & \tiny{\text{\sffamily X} (1.000)}
\\ \hline
\small{Windows} & \tiny{\checkmark} & \tiny{\checkmark} & \tiny{\checkmark} & \tiny{\checkmark} & \tiny{\checkmark} & \tiny{\text{\sffamily X} (0.012)} & \tiny{\checkmark} & \tiny{\checkmark} & \tiny{\checkmark} & \tiny{\checkmark} & \tiny{\checkmark} & \tiny{\checkmark}
\\ \hline
\small{Linux} & \tiny{\checkmark} & \tiny{\checkmark} & \tiny{\checkmark} & \tiny{\checkmark} & \tiny{\checkmark} & \tiny{\text{\sffamily X} (1.000)} & \tiny{\checkmark} & \tiny{\checkmark} & \tiny{\checkmark} & \tiny{\checkmark} & \tiny{\checkmark} & \tiny{\checkmark}
\\ \hline

\end{tabular}}
\caption{Evidence of feature-dependent label noise. A \checkmark indicates that we can reject the null hypothesis $H_0$ that evidence of exploits within a source is independent of the feature. Cells with no p-value are $<0.001$. 
}
\label{table:indep_test_results}
\end{table*}

\topic{Exploit prediction is subject to feature-dependent label noise}
Good predictions also require a judicious solution to the label noise challenge discussed in Section~\ref{sec:challenges}.
The time-varying view of exploitability revealed that our problem is subject to class-dependent noise.
However, because we aggregate evidence about exploits from multiple sources, their individual biases could also affect our ground truth.
To test for such individual biases, we investigate the dependence between all sources of exploit evidence and various vulnerability characteristics.
For each source and feature pair, we perform a Chi-squared test for independence, aiming to observe whether we can reject the null hypothesis $H_0$ that the presence of an exploit within the source is independent of the presence of the feature for the vulnerabilities. 
Table~\ref{table:indep_test_results} lists the results for all 12 sources of ground truth, across the most prevalent vulnerability types and affected products in our dataset.
We utilize the Bonferroni correction for multiple tests~\cite{dunn1961multiple} and a 0.01 significance level.
We observe that the null hypothesis can be rejected for at least 4 features for each source, indicating that \textit{all the sources for ground truth include biases} caused by individual vulnerability features. 
These biases could be reflected in the aggregate ground truth, suggesting that \textit{exploit prediction is subject to class- and feature-dependent label noise}.

\section{Computing Expected Exploitability}
\label{sec:exploit_predictor}


In this section we describe the system for computing \texttt{EE}, starting from the design and implementation of our feature extractor, and presenting the classifier choice.

\begin{table}[t]
\renewcommand{\arraystretch}{1.25}
\begin{center}
\resizebox{\columnwidth}{!}{
\begin{tabular}{ | c | c | c |}
  \hline
\small\strut\textbf{Type} & \small\strut\textbf{Description}  & \small\strut\textbf{\#} \\ \hline
 \multicolumn{3}{|c|}{\textbf{PoC Code (Novel)}} \\ \hline
 \small\strut{Length} 	&	\# characters, loc, sloc	&   33  \\ \hline
 \small\strut{Language} 	&	Programming language label	&   11  \\ \hline
 \small\strut{Keywords count} &  Count for reserved keywords  &   820   \\ \hline
 \small\strut{Tokens} &   Unigrams from code  &   92,485   \\ \hline
  \small{\#\_nodes} 	&	 \# nodes in the AST tree	&	4    \\ \hline
 \small{\#\_internal\_nodes} 	&	 \# of internal AST tree nodes	&   4   \\ \hline
 \small{\#\_leaf\_nodes} 	&	 \# of leaves of AST tree	&   4 \\ \hline
 \small{\#\_identifiers} 	&	 \# of distinct identifiers	&   4   \\ \hline
 \small{\#\_ext\_fun} 	&	 \# of external functions called	&	 4  \\ \hline
 \small{\#\_ext\_fun\_calls} 	&	 \# of calls to external functions	&	 4  \\ \hline
 \small{\#\_udf} 	&	 \# user-defined functions	&	 4  \\ \hline
 \small{\#\_udf\_calls} 	&	 \# calls to user-defined functions	&	 4  \\ \hline
 \small{\#\_operators} 	&	 \# operators used	&	 4  \\ \hline
  \small{cyclomatic compl} 	& cyclomatic complexity	&	 4  \\ \hline
 \small{nodes\_count\_*} 	&	 \# of AST nodes for each node type	&	 316    \\ \hline
 \small{ctrl\_nodes\_count\_*} 	&	 \# of AST nodes for each control statement type	&	 29 \\ \hline
 \small{literal\_types\_count\_*} 	&	 \# of AST nodes for each literal type	&	 6 \\ \hline
 \small{nodes\_depth\_*} 	&	 Stats depth in tree for each AST node type	&	 916    \\ \hline
  \small{branching\_factor} 	&	 Stats \# of children across AST	&	 12  \\ \hline
 \small{branching\_factor\_ctrl} 	&	 Stats \# of children within the Control AST	&	 12  \\ \hline
 \small{nodes\_depth\_ctrl\_*} 	&	 Stats depth in tree for each Control AST node type	&	116  \\ \hline
 \small{operator\_count\_*} 	&	 Usage count for each operator	&	 135 \\ \hline
 \small{\#\_params\_udf} 	&	 Stats \# of parameters for user-defined functions	&	12    \\ \hline
   \multicolumn{3}{|c|}{\textbf{PoC Info (Novel)}} \\ \hline
 \small\strut{PoC unigrams} 	&	 PoCs text and comments	&   289,755   \\ \hline
   \multicolumn{3}{|c|}{\textbf{Write-ups (Prior Work)}} \\ \hline
 \small\strut{Write-up unigrams} 	&	 Write-ups text	&   488,490 	 \\ \hline
 
  \multicolumn{3}{|c|}{\textbf{Vulnerability Info (Prior Work)}} \\ \hline
  \small\strut{NVD unigrams} 	&	 NVD descriptions	&   103,793	\\ \hline
 \small\strut{CVSS} 	&	 CVSSv2 \& CVSSv3 components	&   40  \\ \hline
 \small\strut{CWE} 	&	 Weakness type	&   154 \\ \hline
 \small\strut{CPE} 	&	 Name of affected product	&	10  \\ \hline
   \multicolumn{3}{|c|}{\textbf{In-the-Wild Predictors (Prior Work)}} \\ \hline
 \small\strut{EPSS} 	&	 Handcrafted	&   53 	 \\ \hline
  \small\strut{Social Media} 	&	 Twitter content and statistics	&   898,795	 \\ \hline
\end{tabular}}
\end{center}
\caption{Description of features used. Unigram features are counted before frequency-based pruning. 
}
\label{table:features}
\end{table}

\subsection{Feature Engineering}
\label{sec:feature_engineering}

\texttt{EE} uses features extracted from all vulnerability and PoC artifacts in our datasets, which are summarized in Table~\ref{table:features}.


\topic{Novel: PoC Code} 
Intuitively, one of the leading indicators for the complexity of functional exploits is the complexity of PoCs.
This is because if triggering the vulnerability requires a complex PoC, an exploit would also have to be complex.
Conversely, complex PoCs could already implement functionality beneficial towards the development of functional exploits.
We use this intuition to extract features that reflect the complexity of PoC code, by means of intermediate representations that can capture it.
We transform the code into Abstract Syntax Trees (ASTs), a low-overhead representation which encodes structural characteristics of the code. 
From the ASTs we extract complexity features such as statistics of the node types, structural features of the tree, as well as statistics of control statements within the program and the relationship between them.
Additionally, we extract features for the function calls within the PoCs towards external library functions, which in some cases may be the means through which the exploit interacts with the vulnerability and thereby reflect the relationship between the PoC and its vulnerability.
Therefore, the library functions themselves, as well as the patterns in calls to these functions, can reveal information about the complexity of the vulnerability, which might in turn express the difficulty of creating a functional exploit.
We also extract the cyclomatic complexity from the AST~\cite{landman2016empirical}, a software engineering metric 
which encodes the number of independent code paths in the program.
Finally, we encode features of the PoC programming language, in the form of statistics over the file size and the distribution of language reserved keywords.

We also observe that the lexical characteristics of the PoC code provide insights into the complexity of the PoC.
For example, a variable named \texttt{shellcode} in a PoC might suggest that the exploit is in an advanced stage of development.
In order to capture such characteristics, we extract the code tokens from the entire program, capturing literals, identifiers and reserved keywords, in a set of binary unigram features.
Such specific information allows us to capture the stylistic characteristics of the exploit, the names of the library calls used, as well as more latent indicators, such as artifacts indicating exploit authorship~\cite{caliskan2015anonymizing}, which might provide utility towards predicting exploitability.
Before training the classifier, we filter out lexicon features that appear in less than 10 training-time PoCs, which helps prevent overfitting.

\topic{Novel: PoC Info} Because a large fraction of PoCs contain only textual descriptors for triggering the vulnerabilities without actual code, we also extract features that aim to encode the technical information conveyed by the authors in the non-code PoCs, as well as comments in code PoCs.
We encode these features as binary unigrams.
Unigrams provide a clear baseline for the performance achievable using NLP.
Nevertheless, in 
\arxiv{
Section~\ref{sec:effectiveness-of-exploitability}
}
\usenix{
our technical report~\cite{suciu2021expected}
}
we investigate the performance of \texttt{EE} with embeddings, showing that there are additional challenges in designing semantic NLP features for exploit prediction, which we leave for future work.


\topic{Prior Work: Vulnerability Info  and Write-ups}
To capture the technical information shared through natural language in artifacts, we extract unigram features from all the write-ups discussing each vulnerability and the NVD descriptions of the vulnerability.
Finally, we extract the structured data within NVD that encodes vulnerability characteristics: the most prevalent list of products affected by the vulnerability, the vulnerability types (CWEID~\cite{MITRE:CWE}), and all the CVSS Base Score sub-components, using one-hot encoding.

\topic{Prior Work: In-the-Wild Predictors} 
To compare the effectiveness of various feature sets, we also extract 2 categories proposed in prior predictors of exploitation in the wild.
The Exploit Prediction Scoring System (EPSS)~\cite{jacobs2019exploit} proposes 53 features manually selected by experts as good indicators for exploitation in the wild. This set of \textit{handcrafted features} contains tags reflecting vulnerability types, products and vendors, as well as binary indicators of whether PoC or weaponized exploit code has been published for a vulnerability. 
Second, from our collection of tweets, we extract \textit{social media features} introduced in prior work~\cite{USENIX-Security-2015}, which reflect the textual description of the discourse on Twitter, as well as characteristics of the user base and tweeting volume for each vulnerability. 
Unlike the original work, we do not perform feature selection on the unigram features from tweets because we want to directly compare the utility of Twitter discussions to these from other artifacts.
None of the two categories will be used in the final \texttt{EE} model because of their limited predictive utility.

\subsection{Feature Extraction System}
\label{sec:feature_extraction_system}

\begin{figure}[t]
\centering
\includegraphics[height=1.3in]{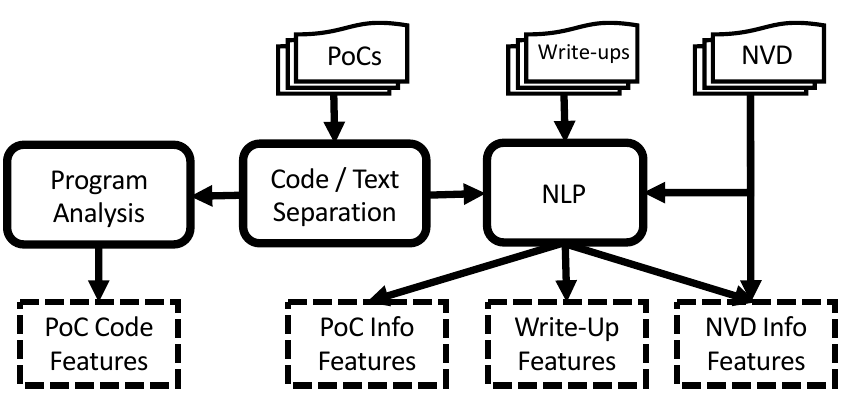}
\caption{Diagram of the \texttt{EE} feature extraction system.}
\label{fig:processing_diagram}
\end{figure}

Below we describe the components of our feature extraction system, illustrated in Figure~\ref{fig:processing_diagram}, and discuss how we address the challenges identified in Section~\ref{sec:challenges}.

\topic{Code/Text Separation} 
Only 64\% of the PoCs in our dataset contain any file extension that would allow us to identify the programming language.
Moreover, 5\% of them have conflicting information from different sources, and we observe that many PoCs are first posted online as freeform text without explicit language information.
Therefore, a central challenge is to accurately identify their programming languages and whether they contain any code.
We use GitHub Linguist~\cite{githublinguist}, to extract the most likely programming languages used in each PoC.                                                              
Linguist combines heuristics with a Bayesian classifier to identify the most prevalent language within a file.
Nevertheless, the model obtains an accuracy of 0.2 on classifying the PoCs, due to the prevalence of natural language text in PoCs.
After modifying the heuristics and retraining the classifier on 42,195 PoCs from ExploitDB that contain file extensions, we boost the accuracy to 0.95.
The main cause of errors is text files with code file extensions, yet these errors have limited impact because of the NLP features extracted from files.

\begin{table}[t]
\renewcommand{\arraystretch}{1.25}
\centering
\begin{tabular}{ | c | c | c |}
  \hline
 \small\textbf{Language} & \small\strut\textbf{\# PoCs}  & \small\strut\textbf{\# CVEs (\% exploited)} \\ \hline
Text & 27743 & 14325 (47\%) \\ \hline
Ruby & 4848 & 1988 (92\%) \\ \hline
C & 4512 & 2034 (30\%) \\ \hline
Perl & 3110 & 1827 (54\%) \\ \hline
Python & 2590 & 1476 (49\%) \\ \hline
JavaScript & 1806 & 1056 (59\%) \\ \hline
PHP & 1040 & 708 (55\%) \\ \hline
HTML & 1031 & 686 (56\%) \\ \hline
Shell & 619 & 304 (29\%) \\ \hline
VisualBasic & 397 & 215 (41\%) \\ \hline
None & 367 & 325 (43\%) \\ \hline
C++ & 314 & 196 (34\%) \\ \hline
Java & 119 & 59 (32\%) \\ \hline
\end{tabular}
\caption{Breakdown of the PoCs in our dataset according to programming language.}
\label{table:pocs_pls}
\end{table}

Table~\ref{table:pocs_pls} lists the number of PoCs in our dataset for each identified language label (the None label represents the cases which our classifier could not identify any language, including less prevalent programming languages not in our label set).
We observe that 58\% of PoCs are identified as text, while the remaining ones are written in a variety of programming languages.
Based on this separation, we develop regular expressions to extract the comments from all code files. 
After separating the comments, we process them along with the text files using NLP, to obtain \textit{PoC Info} features, while the \textit{PoC Code} features are obtained using NLP and program analysis.

\topic{Code Features} 
Performing program analysis on the PoCs poses a challenge because many of them do not have a valid syntax or have missing dependencies that hinders compilation or interpretation~\cite{Allodi12:VulnerabilityScores,mu2018understanding}. 
We are not aware of any unified and robust solution to simultaneously obtain ASTs from code written in different languages.
We address this challenge by employing heuristics to correct malformed PoCs and parsing them into intermediate representations using techniques that provide robustness to errors.

Based on Table~\ref{table:pocs_pls}, we observe that some languages are likely to have a more significant impact on the prediction performance, based on prevalence and frequency of functional exploits among the targeted vulnerabilities.
Given this observation, we focus our implementation on Ruby, C/C++, Perl and Python.
Note that this choice does not impact the extraction of lexical features from code PoCs written in other languages.

For C/C++ we use the Joern fuzzy parser for program analysis, previously proposed for bug discovery~\cite{yamaguchi2014modeling}.
The tool provides robustness to parsing errors through the use of island grammars and allows us to successfully parse 98\% of the files.

On Perl, by modifying the existing \emph{Compiler::Parser}~\cite{perlparser} tool to improve its robustness, and employing heuristics to correct malformed PoC files, we improve the parsing success rate from 37\% to 83\%. 

For Python, we implement a feature extractor based on the \emph{ast} parsing library\cite{pythonparser}, achieving a success rate of 67\%. 
We observe that this lower parsing success rate is due to the reliance of the language on strict indentation, which is often distorted or completely lost when code gets distributed through Webpages.

Ruby provides an interesting case study because, despite being the most prevalent language among PoCs, it is also the most indicative of exploitation.
We observe that this is because our dataset contains functional exploits from the Metasploit framework, which are written in Ruby.
We extract AST features for the language using the Ripper library~\cite{rubyparser}. 
Our implementation is able to successfully parse 96\% of the files. 

Overall, we successfully parse 13,704 PoCs associated with 78\% of the CVEs that have PoCs with code. 
Each vulnerability aggregates only the code complexity features of the most complex PoC (in source lines of code) across each of the four languages, while the remaining code features are collected from all PoCs available.

\topic{Unigram Features} 
We extract the textual features using a standard NLP pipeline which involves tokenizing the text from the PoCs or vulnerability reports, removing non-alphanumeric characters, filtering out English stopwords and representing them as unigrams.
For each vulnerability, the PoC unigrams are aggregated across all PoCs, and separately across all write-ups collected within the observation period. 
When training a classifier, we discard unigrams which occur less than 100 times across the training set, because they are unlikely to generalize over time and we did not observe any noticeable performance boost when including them.

%

\subsection{Exploit Predictor Design}
\label{sec:design_exploit_predictor}

The predictor concatenates all the extracted features, and uses the ground truth about exploit evidence, to train a classifier which outputs the \texttt{EE} score.
The classifier uses a feedforward neural network having 2 hidden layers of size 500 and 100 respectively, with ReLU activation functions. 
This choice was dictated by two main characteristics of our domain: feature dimensionality and concept drift.
First, as we have many potentially useful features, but with limited coverage, linear models, such as SVM, which tend to emphasize few important features~\cite{melis2018explaining}, would perform worse.
Second, deep learning models are believed to be more robust to concept drift and the shifting utility of features~\cite{pendlebury2019tesseract}, which is a prevalent issue in the exploit prediction task~\cite{USENIX-Security-2015}.
The architecture was chosen empirically by measuring performance for various settings.

\topic{Classifier training}
To address the second challenge identified in Section~\ref{sec:challenges}, we incorporate noise robustness in our system by exploring several loss functions for the classifier.
Our design choices are driven by two main requirements: (i) providing robustness to both class- and feature- dependent noise, and (ii) providing minimal performance degradation when noise specification is not available. 

\setlength{\abovedisplayskip}{2pt}
\setlength{\belowdisplayskip}{2pt}
\textbf{BCE:} The binary cross-entropy is the standard, noise-agnostic loss for training binary classifiers. For a set of $N$ examples $x_i$ with labels $y_i \in \{0,1\}$, the loss is computed as: 
$$L_{BCE}=-\frac{1}{N}\sum\limits_{i=1}^N[y_ilog(p_\theta(x_i))+(1-y_i)log(1-p_\theta(x_i))]$$ 
where $p_\theta(x_i)$ corresponds to the output probability predicted by our classifier.
BCE does not explicitly address requirement (i), but can be used to benchmark noise-aware losses that aim to address (ii).  

\textbf{LR:} The Label Regularization, initially proposed as a semi-supervised loss to learn from unlabeled data~\cite{mann2007simple}, has been shown to address class-dependent label noise in malware classification~\cite{deloach2016android} using a logistic regression classifier.
$$L_{LR}=-\frac{1}{N}\sum\limits_{i=1}^N[y_ilog(p_\theta(x_i))] - \lambda KL(\widetilde{p}||\hat{p}_\theta)$$ 
The loss function complements the log-likelihood loss over the positive examples with a label regularizer, which is the KL divergence between a noise prior $\widetilde{p}$ and the classifier's output distribution over the negative examples $\hat{p}_\theta$:
$$\hat{p}_\theta = \frac{1}{N}\sum\limits_{i=1}^N[(1-y_i)log(1-p_\theta(x_i))]$$
Intuitively, the label regularizer aims to push the classifier predictions on the noisy class towards the expected noise prior $\widetilde{p}$, while the $\lambda$ hyper-parameter controls the regularization strength. 
We use this loss to observe the extent to which existing noise correction approaches for related security tasks apply to our problem.
However, this function was not designed to address (ii) and, as our results will reveal, yields poor performance in our problem.

\textbf{FC:} The Forward Correction loss has been shown to significantly improve robustness to class-dependent label noise in various computer vision tasks~\cite{patrini2017making}.
The loss requires a pre-defined noise transition matrix $T \in [0,1]^{2x2}$, where each element represents the probability of observing a noisy label $\widetilde{y}_j$ for a true label $y_i$: $T_{ij}=p(\widetilde{y}_j|y_i)$.
For an instance $x_i$, the log-likelihood is then defined as $l_c(x_i) = -log (T_{0c}(1-p_\theta(x_i)) +  T_{1c}p_\theta(x_i))$ for each class $c \in \{0,1\}$.
In our case, because we assume that the probability of falsely labeling non-exploited vulnerabilities as exploited is negligible, the noise matrix can be defined as:
$T = 
\begin{pmatrix}
  1 & 0\\ 
  \widetilde{p} & 1-\widetilde{p}
\end{pmatrix}
$, and the loss reduces to:
$$
L_{FC}=-\frac{1}{N}\sum\limits_{i=1}^N[y_ilog((1-\widetilde{p})p_\theta(x_i))+
$$
$$+ (1-y_i)log(1-(1-\widetilde{p})p_\theta(x_i))]
$$

On the negative class, the loss reduces the penalty for confident positive predictions, allowing the classifier to output a higher score for predictions which might have noisy labels.
This prevents the classifier from fitting of instances with potentially noisy labels. 
\usenix{
We analyze the loss in more detail in the technical report~\cite{suciu2021expected}.
}
\arxiv{

\begin{figure}[t]
    \begin{subfigure}{.24\textwidth}
        \centering
        \includegraphics[height=1.7in]{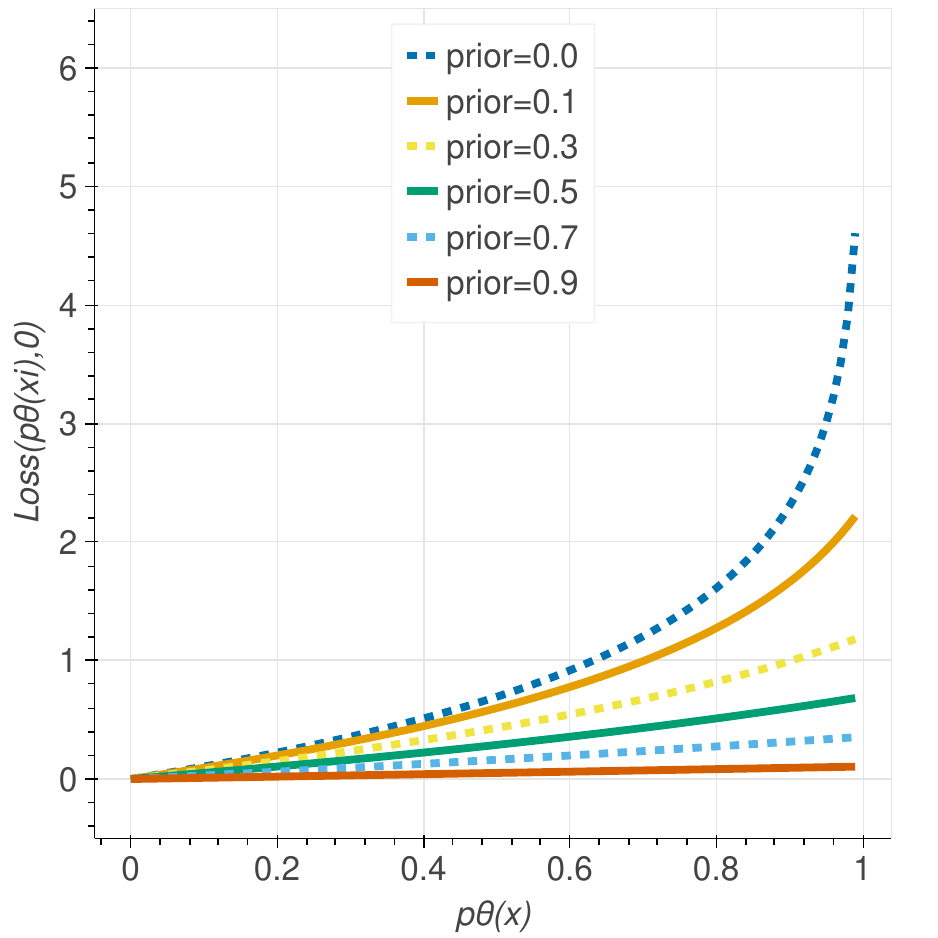}
        \vspace{-0.2in}
        \caption{}
        \label{fig:fcclossa}
    \end{subfigure}%
    \begin{subfigure}{.24\textwidth}
        \centering
        \includegraphics[height=1.7in]{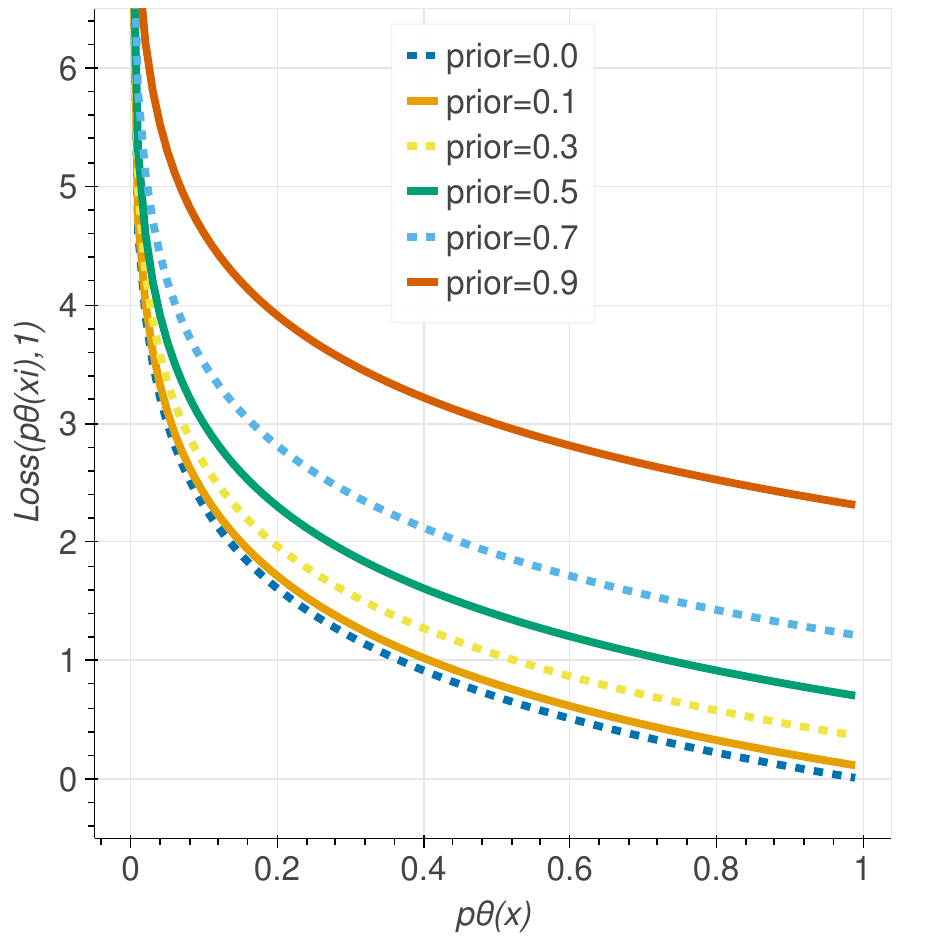}
        \vspace{-0.2in}
        \caption{}
        \label{fig:fcclossb}
    \end{subfigure}%
    \caption{ Value of the FC loss function of the output $p_\theta(x_i)$, for different levels of prior $\widetilde{p}$, when $y=0$ (a) and $y=1$ (b) }
    \label{fig:fccloss}
\end{figure}

Figure~\ref{fig:fccloss} plots the value of the loss function on a single example, for both classes and across the range of priors $\widetilde{p}$. On the negative class, the loss reduces the penalty for confident positive predictions, allowing the classifier to output a higher score for predictions which might have noisy labels. 
This prevents the classifier from fitting of instances with potentially noisy labels. 
}
FC partially addresses requirement (i), being explicitly designed only for class-dependent noise.
However, unlike LR, it naturally addresses (ii) because it is equivalent to BCE if $\widetilde{p}=0$.

\textbf{FFC:} To fully address (i), we modify FC to account for feature-dependent noise, a loss function we denote \textit{Feature Forward Correction (FFC)}. 
We observe that for exploit prediction, feature-dependent noise occurs within the same label flipping template as class-dependent noise.
We use this observation to expand the noise transition matrix with instance specific priors: $T_{ij}(x)=p(\widetilde{y}_j|x,y_i)$.
In this case the transition matrix becomes: 
$$T(x) = 
\begin{pmatrix}
  1 & 0\\ 
  \widetilde{p}(x) & 1-\widetilde{p}(x)
\end{pmatrix}
$$
Assuming that we only possess priors for instances that have certain features $f$, the instance prior can be encoded as a lookup-table:
$$
\widetilde{p}(x,y) = \begin{cases} \widetilde{p}_f &\mbox{if }  y = 0\mbox{ and }x \mbox{ has } f \\
0 & \mbox{otherwise} \end{cases}
$$
While feature-dependent noise might cause the classifier to learn a spurious correlation between certain features and the wrong negative label, this formulation mitigates the issue by reducing the loss only on the instances that possess these features.
In Section~\ref{sec:results} we show that feature-specific prior estimates are achievable from a small set of instances, and use this observation to compare the utility of class- and feature-specific noise priors in addressing label noise.
When training the classifier, we discovered optimal performance when using an ADAM optimizer for 20 epochs and a batch size of 128, using a learning rate of 5e-6.

\topic{Classifier deployment}
We evaluate the historic performance of our classifier by partitioning the dataset into temporal splits, assuming that the classifier is re-trained periodically, on all the historical data available at that time. 
At the time the classifier is trained, we do not include the vulnerabilities disclosed within the last year because the positive labels from exploitation evidence might not be available until later on.
We discovered that the classifier needs to be retrained every six months, as less frequent retraining would affect performance due to a larger time delay between the disclosure of training and testing instances.
During testing, the system operates in a streaming environment in which it continuously collects the data published about vulnerabilities then recomputes their feature vectors over time and predicts their updated \texttt{EE} score.
The prediction for each test-time instance is performed with the most recently trained classifier.
To observe how our classifier performs over time, we train the classifier using the various loss functions and test its performance on all vulnerabilities disclosed between January 2010, when 65\% of our dataset was available for training, and March 2020.

\section{Evaluation}
\label{sec:results}

We evaluate our approach of predicting expected exploitability by testing \ee{} on real-world vulnerabilities and answering the following questions, which are designed to address the third challenge identified in Section~\ref{sec:challenges}:
How effective is \ee{} at addressing label noise?
How well does \ee{} perform compared to baselines?
How well do various artifacts predict exploitability?
How does \ee{} performance evolve over time?
Can \ee{} anticipate imminent exploits?
Does \ee{} have practicality for vulnerability prioritization?

\begin{table}[t]
\renewcommand{\arraystretch}{1.25}
\setlength{\tabcolsep}{4.2pt}
\centering
\resizebox{0.9\columnwidth}{!}{
\scriptsize
\begin{tabular}{ | c | c | c | c | c |}
  \hline
\scriptsize\strut\textbf{Feature} & \scriptsize\textbf{\% Noise} & \scriptsize\strut\textbf{Actual Prior} & \scriptsize\strut\textbf{Est. Prior} & \scriptsize\strut\textbf{\# Inst to Est.} \\ \hline
\scriptsize{CWE-79} & 14\% & 0.93 & 0.90 & 29 \\ \hline 
\scriptsize{CWE-94} & 7\% & 0.36 & 0.20 & 5 \\ \hline 
\scriptsize{CWE-89} & 20\% & 0.95 & 0.95 & 22 \\ \hline 
\scriptsize{CWE-119} & 14\% & 0.44 & 0.57 &  51 \\ \hline 
\scriptsize{CWE-20} & 6\% & 0.39 & 0.58 & 26 \\ \hline 
\scriptsize{CWE-22} & 8\% & 0.39 & 0.80 & 15 \\ \hline 
\scriptsize{Windows} & 8\% & 0.35 & 0.87 & 15 \\ \hline 
\scriptsize{Linux}& 5\% & 0.32 & 0.50 & 4 \\ \hline 
\end{tabular}}
\caption{Noise simulation setup. We report the \% of negative instances that are noisy, the actual and estimated noise prior, and the \# of instances used to estimate the prior.}
\label{table:noise_setup}
\end{table}

\begin{table}[t]
\setlength{\tabcolsep}{5pt}
\renewcommand{\arraystretch}{1.25}
\centering
\resizebox{\columnwidth}{!}{
\begin{tabular}{ | c | c | c || c | c || c | c || c | c |}
  \hline
 & \multicolumn{2}{c||}{\small\strut\textbf{BCE}} & \multicolumn{2}{c||}{\small\strut\textbf{LR}} & \multicolumn{2}{c||}{\small\strut\textbf{FC}} & \multicolumn{2}{c|}{\small\strut\textbf{FFC}} \\ \hline
\scriptsize{Feature} & \scriptsize{P} & \scriptsize{AUC} & \scriptsize{P} & \scriptsize{AUC} & \scriptsize{P} & \scriptsize{AUC} & \scriptsize{P} & \scriptsize{AUC} \\ \hline
\scriptsize{CWE-79} & 0.58 & 0.80 & 0.67 & 0.79 & 0.58 & 0.81 & 0.75 & 0.87 \\ \hline 
\scriptsize{CWE-94} & 0.81 & 0.89 & 0.71 & 0.81 & 0.81 & 0.89 & 0.82 & 0.89 \\ \hline 
\scriptsize{CWE-89} & 0.61 & 0.82 & 0.57 & 0.74 & 0.61 & 0.82 & 0.81 & 0.89 \\ \hline 
\scriptsize{CWE-119} & 0.78 & 0.88 & 0.75 & 0.83 & 0.78 & 0.87 & 0.81 & 0.89 \\ \hline 
\scriptsize{CWE-20} & 0.81 & 0.89 & 0.72 & 0.82 & 0.80 & 0.88 & 0.82 & 0.90 \\ \hline 
\scriptsize{CWE-22} & 0.81 & 0.89 & 0.69 & 0.80 & 0.81 & 0.89 & 0.83 & 0.90 \\ \hline 
\scriptsize{Windows} & 0.80 & 0.88 & 0.71 & 0.81 & 0.80 & 0.88 & 0.83 & 0.90 \\ \hline 
\scriptsize{Linux} & 0.81 & 0.89 & 0.71 & 0.81 & 0.81 & 0.89 & 0.82 & 0.90 \\ \hline 
\end{tabular}}
\caption{Noise simulation results. We report the precision at a 0.8 recall (P) and the precision-recall AUC. The pristine BCE classifier performance is 0.83 and 0.90 respectively.}
\label{table:noise_results}
\end{table}

\subsection{Feature-dependent Noise Remediation}
\label{sec:feature-dependent-noise-remediation}
To observe the potential effect of feature-dependent label noise on our classifier, we simulate a worst-case scenario in which our training-time ground truth is missing all the exploits for certain features.
The simulation involves training the classifier on dataset \texttt{DS2}, on a ground truth where all the vulnerabilities with a specific feature $f$ are considered not exploited.
At testing time, we evaluate the classifier on the original ground truth labels. 
Table~\ref{table:noise_setup} describes the setup for our experiments. 
We investigate 8 vulnerability features, part of the Vulnerability Info category, that we analyzed in Section~\ref{sec:empirical_results}: the six most prevalent vulnerability types, reflected through the CWE-IDs, as well as the two most popular products: \texttt{linux} and \texttt{windows}.
Mislabeling instances with these features results in a wide range of noise: between 5-20\% of negative labels become noisy during training.

All techniques require priors about the probability of noise.
The LR and FC approaches require a prior $\widetilde{p}$ over the entire negative class.
To evaluate an upper bound of their capabilities, we assume perfect prior and set $\widetilde{p}$ to match the fraction of training-time instances that are mislabeled.
The FFC approach assumes knowledge of the noisy feature $f$.
This assumption is realistic, as it is often possible to enumerate the features that are most likely noisy (e.g. prior work identified \texttt{linux} as a noise-inducing feature due to the fact that the vendor collecting exploit evidence does not have a product for the platform~\cite{USENIX-Security-2015}).
Besides, FFC requires estimates of the feature-specific priors $\widetilde{p}_f$.
We assume an operational scenario were $\widetilde{p}_f$ is estimated once, by manually labeling a subset of instances collected after training.
We use the vulnerabilities disclosed in the first 6 months after training for estimating $\widetilde{p}_f$ and require that these vulnerabilities are correctly labeled.
Table~\ref{table:noise_setup} shows the actual and the estimated priors $\widetilde{p}_f$, as well as the number of instances used for the estimation.
We observe that the number of instances required for estimation is small, ranging from 5 to 51 across all features $f$, which demonstrates that setting feature-based priors is feasible in practice.
Nevertheless, we observe that the estimated priors are not always accurate approximations of the actual ones, which might negatively impact FFC's ability to address the effect of noise.

In Table~\ref{table:noise_results} we list the results of our experiment.
For each classifier, we report the precision achievable at a recall of 0.8, as well as the precision-recall AUC.
Our first observation is that the performance of the vanilla BCE classifier is not equally affected by noise across different features.
Interestingly, we observe that the performance drop does not appear to be linearly dependent on the amount of noise: both \texttt{CWE-79} and \texttt{CWE-119} result in 14\% of the instances being poisoned, yet only the former inflicts a substantial performance drop on the classifier.
Overall, we observe that the majority of the features do not result in significant performance drops, suggesting that BCE offers a certain amount of built-in robustness 
to feature-dependent noise, possibly due to redundancies in the feature space which cancel out the effect of the noise.

For LR, after performing a grid search for the optimal $\lambda$ parameter which we set to 1, we were unable to match the BCE performance on the pristine classifier.
Indeed, we observe that the loss is unable to correct the effect of noise on any of the features, suggesting that it is not a suitable choice for our classifier as it does not address any of the two requirements of our classifier.

On features where BCE is not substantially affected by noise, we observe that FC performs similarly well.
However, on \texttt{CWE-79} and \texttt{CWE-89}, the two features which inflict the most performance drop, FC is not able to correct the noise even with perfect priors, highlighting the inability of the existing technique to capture feature-dependent noise.
In contrast, we observe that FFC provides a significant performance improvement.
Even for the feature inducing the most degradation, \texttt{CWE-79}, the FFC AUC is restored within 0.03 points of the pristine classifier, although suffering a slight precision drop.
On most features, FCC approaches the performance of the pristine classifier, in spite of being based on inaccurate prior estimates. 

Our result highlights the overall benefits of identifying potential sources of feature-dependent noise, as well as the need for noise correction techniques tailored to our problem.
In the remainder of this section, we will use the FFC with $\widetilde{p}_f=0$ (which is equivalent to BCE), in order to observe how the classifier performs in absence of any noise priors.


\begin{figure}[t]
    \begin{subfigure}{.24\textwidth}
        \centering
        \includegraphics[height=1.8in]{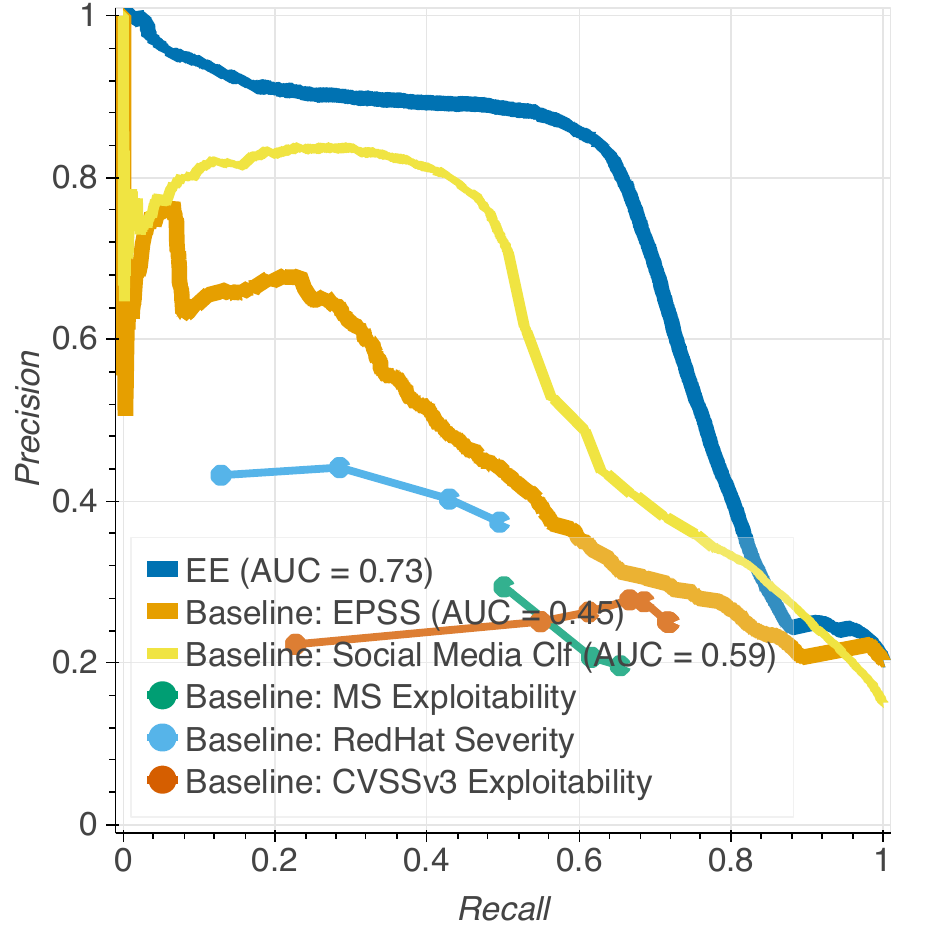}
        \vspace{-0.2in}
        \caption{}
        \label{fig:prediction_performance1a}
    \end{subfigure}%
    \begin{subfigure}{.24\textwidth}
        \centering
        \includegraphics[height=1.8in]{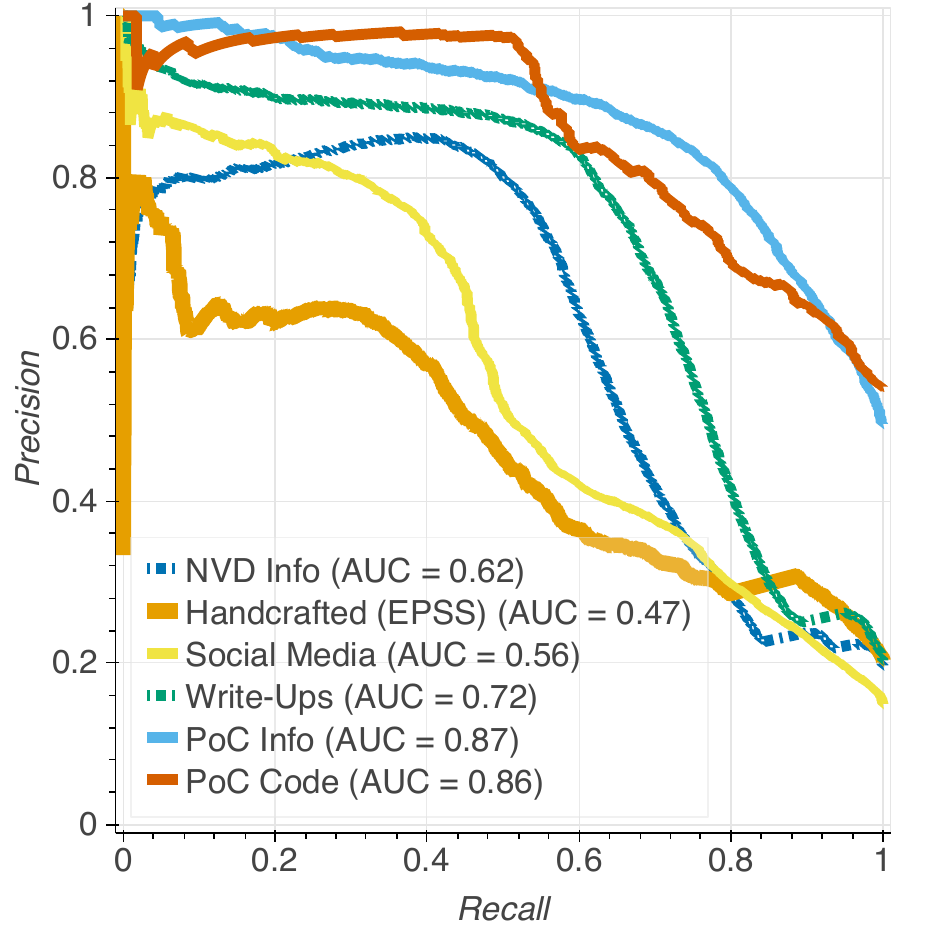}
        \vspace{-0.2in}
        \caption{}
        \label{fig:prediction_performance1b}
    \end{subfigure}%
    \caption{ Performance, evaluated 30 days after disclosure, of (a) \texttt{EE} compared to baselines, (b) individual feature categories. We report the Area under the Curve (AUC) and list the corresponding TPR/FPR curves in Appendix~\ref{appendix:a1}.}
    \label{fig:prediction_performance1}
    
\end{figure}

\subsection{Effectiveness of Exploitability Prediction}
\label{sec:effectiveness-of-exploitability}
Next, we evaluate the effectiveness of our system compared to the three static metrics described in Section~\ref{sec:method}, as well as two state-of-the-art classifiers from prior work.
These two predictors, EPSS~\cite{jacobs2019exploit}, and the Social Media Classifier (SMC)~\cite{USENIX-Security-2015}, were proposed for exploits in the wild and we re-implement and re-train them for our task.
EPSS trains an ElasticNet regression model on the set of 53 hand-crafted features extracted from vulnerability descriptors.
SMC combines the social media features with vulnerability information features from NVD to learn a linear SVM classifier.
For both baselines, we perform hyper-parameter tunning and report the highest performance across all experiments, obtained using $\lambda=0.001$ for EPSS and $C=0.0001$ for SMC.
SMC is trained starting from 2015, as our tweets collection does not begin earlier.

In Figure~\ref{fig:prediction_performance1a} we plot the precision-recall trade-off of the classifiers trained on dataset \texttt{DS1}, evaluated 30 days after the disclosure of test-time instances.
We observe that none of the static exploitability metrics exceed 0.5 precision, while \texttt{EE} significantly outperforms all the baselines.
The performance gap is especially apparent for the 60\% of exploited vulnerabilities, where \texttt{EE} achieves 86\% precision, whereas the SMC, the second-best performing classifier, obtains only 49\%. 
We also observe that for around 10\% of vulnerabilities, the artifacts available within 30 days have limited predictive utility, which affects the performance of these classifiers.

\topic{\texttt{EE} uses the most informative features}
To understand why \texttt{EE} is able to outperform these baselines, 
in Figure~\ref{fig:prediction_performance1b} we plot the performance of \texttt{EE} trained and evaluated on individual categories of features (i.e., only considering instances which have artifacts within these categories).
We observe that \textit{the handcrafted features are the worst performing category}, perhaps due to the fact that the 53 features are not sufficient to capture the large diversity of vulnerabilities in our dataset.
These features encode the existence of public PoCs, which is often used by practitioners as a heuristic rule for determining which vulnerabilities must be patched urgently. 
Our results suggest that this heuristic provides a weak signal for the emergence of functional exploits, in line with prior 
work predicting exploits~\cite{Allodi12:VulnerabilityScores,jacobs2019exploit,Tavabi2018DarkEmbedEP}, which concluded that PoCs "are not a reliable source of information for exploits in the wild"~\cite{Allodi12:VulnerabilityScores}.
Nevertheless, we can achieve a much higher precision at predicting exploitability by extracting deeper features from the PoCs. 
The PoC Code features provide a 0.93 precision for half of the exploited vulnerabilities, outperforming all other categories.
This suggests that \textit{code complexity can be a good indicator for the likelihood of functional exploits}, although not on all instances, as indicated by the sharp drop in precision beyond the 0.5 recall.
A major reason for this drop is the existence of post-exploit mitigation techniques: even if a PoC is complex and contains advanced functionality, defenses might impede successful exploitation beyond denial of service.
This highlights how our feature extractor is able to represent PoC descriptions and code characteristics which reflect exploitation efforts.
Both the PoC and Write-up features, which \texttt{EE} capitalizes on, perform significantly better than other categories.

Surprisingly, we observe that social media features, are not as useful for predicting functional exploits as they are for exploits in the wild~\cite{USENIX-Security-2015}, a finding reinforced by 
\usenix{
our additional experiments from the technical report~\cite{suciu2021expected},
}
\arxiv{
the results of the experiments conducted below, 
}
which show that they do not improve upon other categories. 
This is because tweets tend to only summarize and repeat information from write-ups, and often do not contain sufficient technical information to predict exploit development.
Besides, they often incur an additional publication delay over the original write-ups they quote.
Overall, our evaluation highlights a qualitative distinction between the problem of predicting functional exploits and that of predicting exploits in the wild. 

\begin{figure}[t]
    \begin{subfigure}{.24\textwidth}
        \centering
        \includegraphics[height=1.8in]{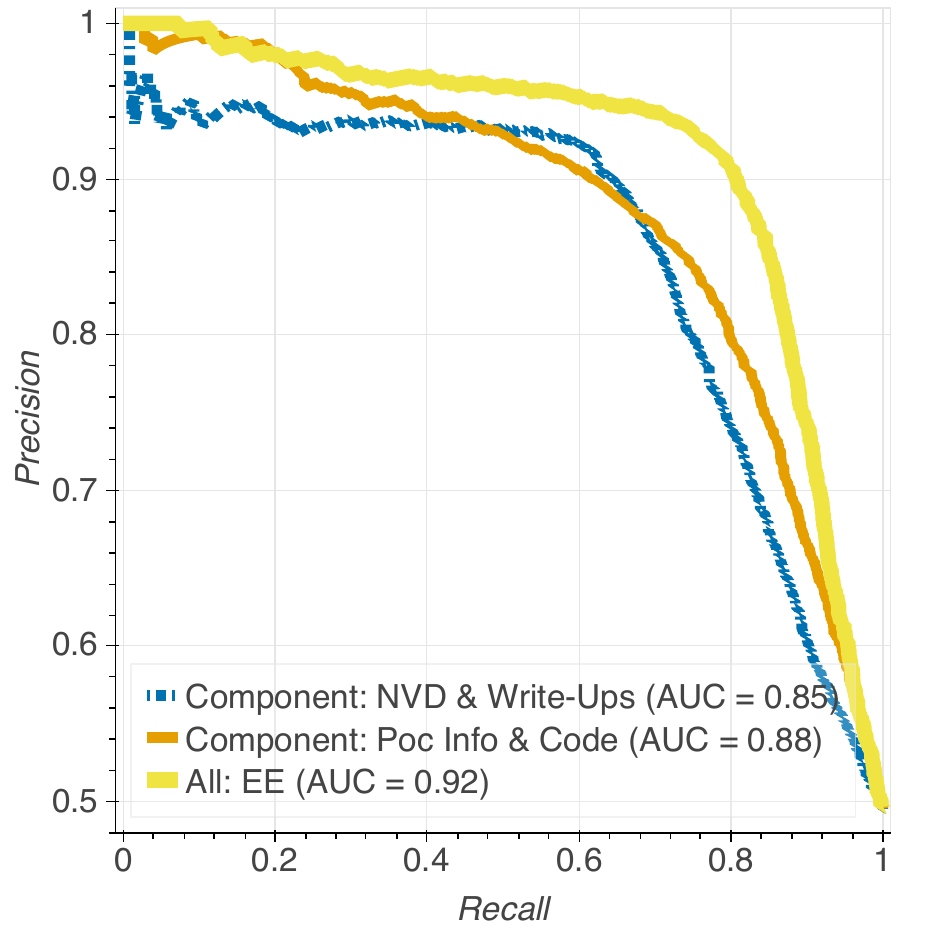}
        \vspace{-0.2in}
        \caption{}
        \label{fig:prediction_performance2a}
    \end{subfigure}%
    \begin{subfigure}{.24\textwidth}
        \centering
        \includegraphics[height=1.8in]{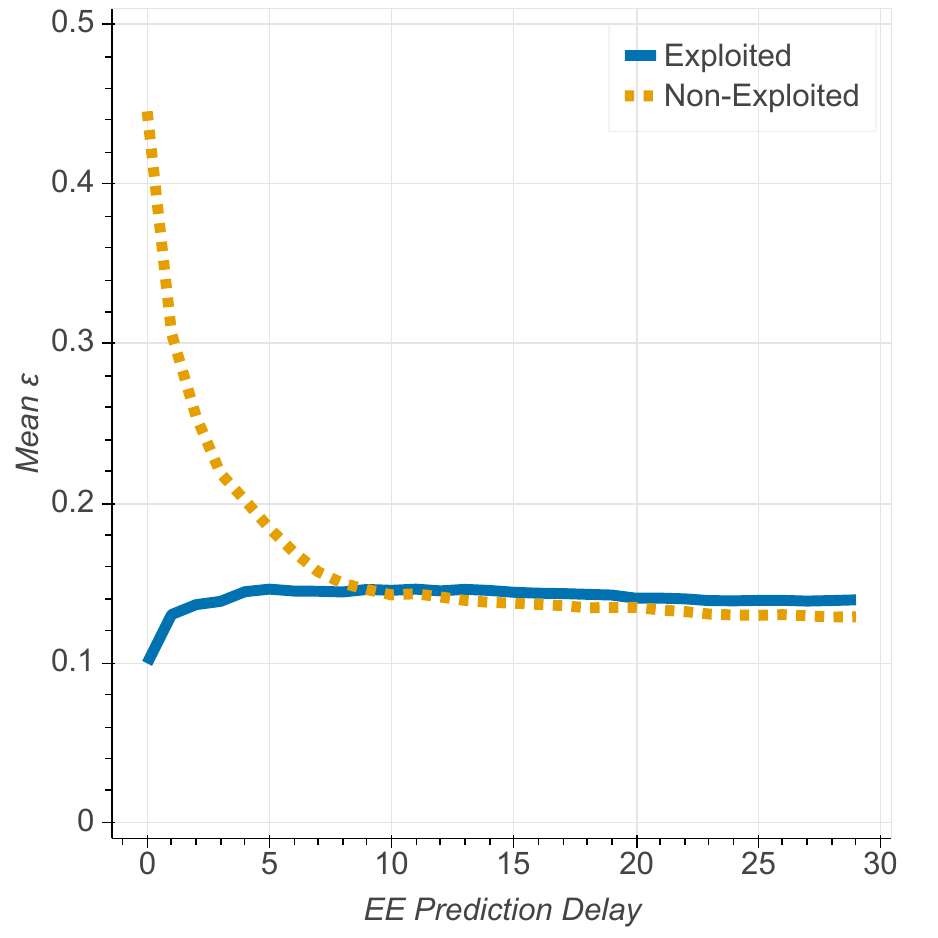}
        \vspace{-0.2in}
        \caption{}
        \label{fig:prediction_performance2b}
    \end{subfigure}%
    \caption{(a) Performance of \texttt{EE} compared to constituent subsets of features. (b) $\mathcal{P}$ evaluated at different points in time.}
    \label{fig:prediction_performance2}
\vspace{-10pt}
\end{figure}

\topic{\texttt{EE} improves when combining artifacts}
Next, we look at the interaction among features on dataset \texttt{DS2}.
In Figure~\ref{fig:prediction_performance2a} we compare the performance of \texttt{EE} trained on all feature sets, with that trained on PoCs and vulnerability features alone.
We observe that PoC features outperform these from vulnerabilities, while their combination results in a significant performance improvement.
The result highlights the two categories complement each other and confirm that PoC features provide additional utility for predicting exploitability.
On the other hand, 
\usenix{
as described in detail in the technical report~\cite{suciu2021expected},
}
\arxiv{
as described below, 
}
we observe no added benefit when incorporating social media features into \texttt{EE}. 
We therefore exclude them from our final \texttt{EE} feature set.


\topic{\texttt{EE} performance improves over time}
In order to evaluate the benefits of time-varying exploitability, the precision-recall curves are not sufficient, because they only capture a snapshot of the scores in time. In practice, the \texttt{EE} score would be compared to that of other vulnerabilities disclosed within a short time, based on their most recent scores.
Therefore, we introduce a metric $\mathcal{P}$ to compute the performance of \texttt{EE} in terms of the expected probability of error over time. 

For a given vulnerability $i$, its score $EE_i(z)$ computed on date $z$ and its label $D_{i}$ ($D_{i}=1$ if $i$ is exploited and 0 otherwise), the error $\mathcal{P}^{EE}(z,i,S)$ w.r.t. a set of vulnerabilities $S$ is computed as:
$$
\mathcal{P}^{EE}(z,i,S) = \begin{cases} \frac{||\{D_{j}=0 \land EE_j(z) \geq EE_i(z) | j \in S\}||}{||S||} &\mbox{if } D_{i}=1 \\
\frac{||\{D_{j}=1 \land EE_j(z) \leq EE_i(z) | j \in S\}||}{||S||} &\mbox{if } D_{i}=0 \end{cases}
$$
If $i$ is exploited, the metric reflects the number of vulnerabilities in $S$ which are not exploited but are scored higher than $i$ on date $z$. Conversely, if $i$ is not exploited, $\mathcal{P}$ computes the fraction of exploited vulnerabilities in $S$ which are scored lower than it.
The metric captures the amount of effort spent prioritizing vulnerabilities with no known exploits. For both cases, a perfect score would be 0.0. 

For each vulnerability, we set $S$ to include all other vulnerabilities disclosed within $t$ days after its disclosure. Figure~\ref{fig:prediction_performance2b} plots the mean $\mathcal{P}$ over the entire dataset, when varying $t$ between 0 and 30, for both exploited and non-exploited vulnerabilities.
We observe that on the day of disclosure, \texttt{EE} already provides a high performance for exploited vulnerabilities: on average, only 10\% of the non-exploited vulnerabilities disclosed on the same day will be scored higher than an exploited one. However, the score tends to overestimate the exploitability of non-exploited vulnerabilities, resulting in many false positives. This is in line with prior observations that static exploitability estimates available at disclosure have low precision~\cite{USENIX-Security-2015}. 
By following the two curves along the X-axis, we observe the benefits of time-varying features. 
Over time, the errors made on non-exploited vulnerabilities decrease substantially: while such a vulnerability is expected to be ranked above 44\% exploited ones on the day of disclosure, it will be placed above 14\% such vulnerabilities 10 days later. 
The plot also shows that this sharp performance boost for the non-exploited vulnerabilities incurs a smaller increase in error rates for the exploited class. 
We do not observe great performance improvements after 10 days from disclosure.
Overall, we observe that time-varying exploitability contributes to a \textit{substantial decrease in the number of false positives, therefore improving the precision our estimates}.
To complement our evaluation, the precision-recall trade-offs at various points in time is reported in Appendix~\ref{appendix:a1}.

\arxiv{

\begin{figure}[t]
    \begin{subfigure}{.24\textwidth}
        \centering
        \includegraphics[height=1.8in]{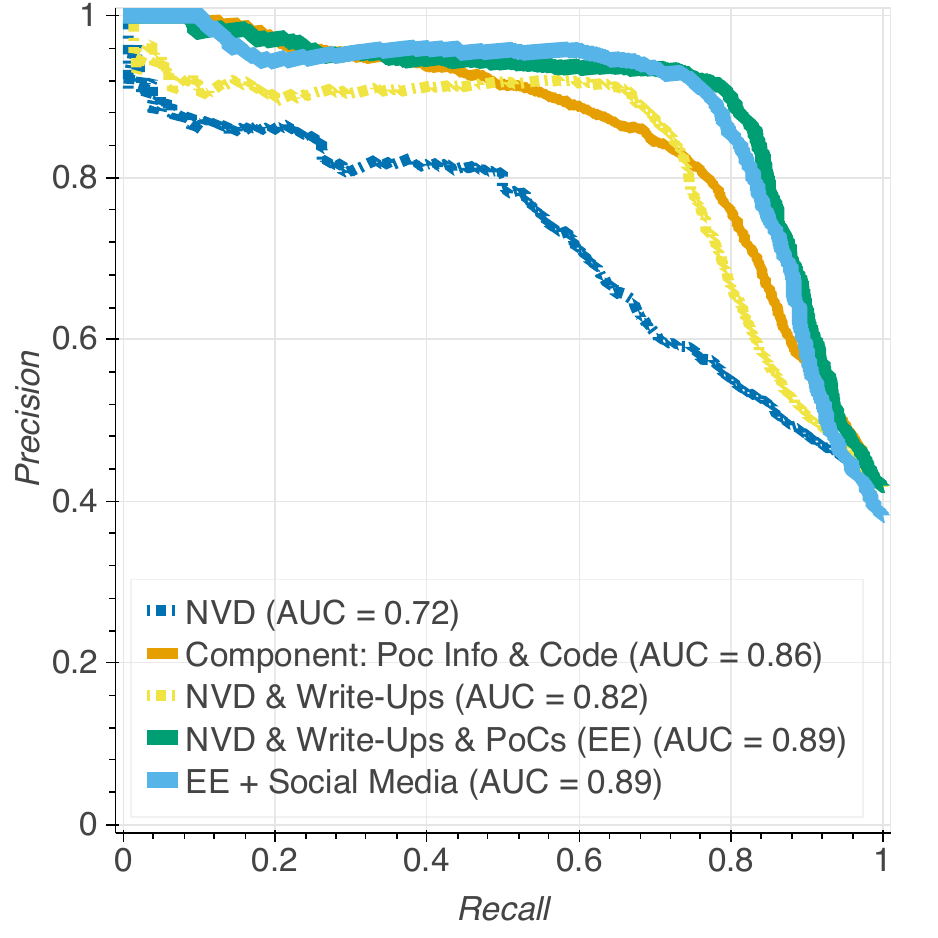}
        \vspace{-0.2in}
        \caption{}
        \label{fig:perf_e_main}
    \end{subfigure}%
    \begin{subfigure}{.24\textwidth}
        \centering
        \includegraphics[height=1.8in]{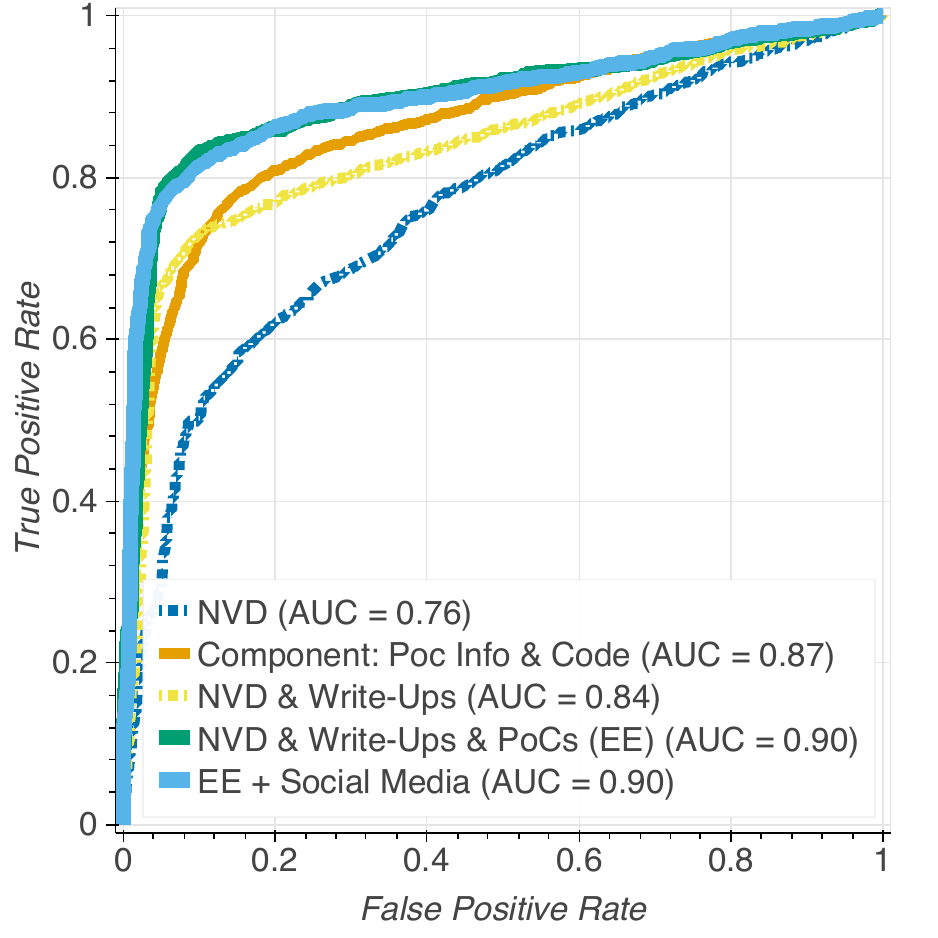}
        \vspace{-0.2in}
        \caption{}
        \label{fig:perf_e_appendix}
    \end{subfigure}%
    \caption{Performance of the classifier when adding Social Media features.}
    \label{fig:perf_e}
\end{figure}

\begin{figure}[t]
    \begin{subfigure}{.24\textwidth}
        \centering
        \includegraphics[height=1.8in]{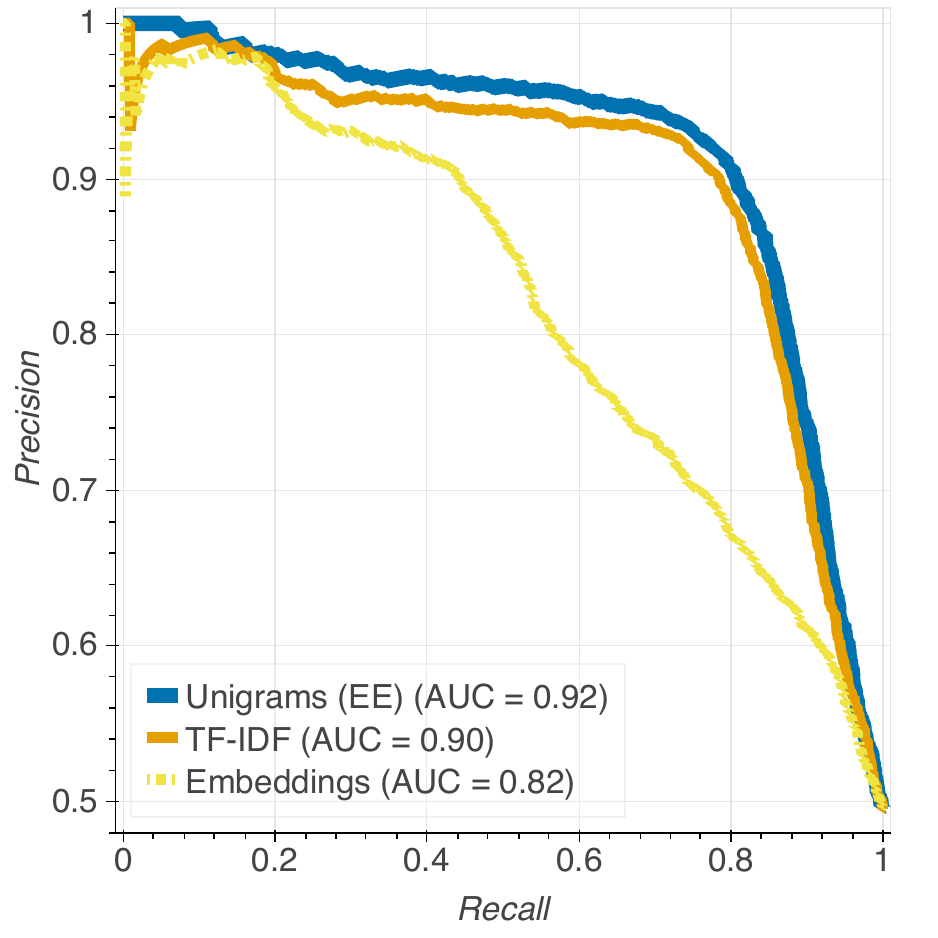}
        \vspace{-0.2in}
        \caption{}
        \label{fig:perf_f_main}
    \end{subfigure}%
    \begin{subfigure}{.24\textwidth}
        \centering
        \includegraphics[height=1.8in]{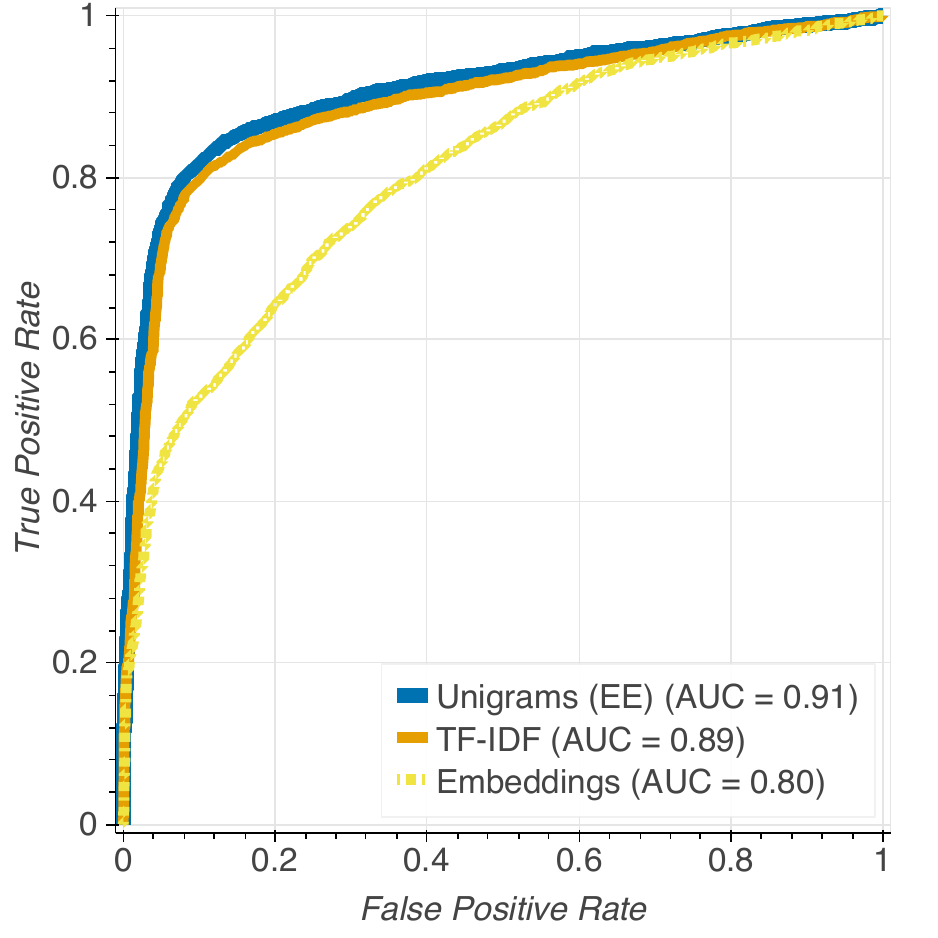}
        \vspace{-0.2in}
        \caption{}
        \label{fig:perf_f_appendix}
    \end{subfigure}%
    \caption{Performance of the classifier when considering additional NLP features.}
    \label{fig:perf_f}
\end{figure}

\topic{Social Media features do not improve \texttt{EE}} 
In Figure~\ref{fig:perf_e} we investigate the effect of adding Social Media features on \texttt{EE}.
The results evaluate the classifier trained on \texttt{DS2}, over the time period spanning our tweets collection.  
We observe that, unlike the addition of PoC features to these extracted from vulnerability artifacts, these features do not improve the performance of the classifier.
This is because tweets generally replicate and summarize the information already contained in the technical write-ups that they link to.
Because these features convey little extra technical information beyond other artifacts, potentially also incurring an additional publication delay, we do not incorporate these features in the final feature set of \texttt{EE}.

\topic{Effect of higher-level NLP features on \texttt{EE}} 
We investigate two alternative representations for natural language features: TF-IDF and paragraph embeddings.
TF-IDF is a common data mining metric used to encode the importance of individual terms within a document, by means of their frequency within a document, scaled by their inverse prevalence across the dataset. 
Paragraph embeddings, which were also used by DarkEmbed~\cite{Tavabi2018DarkEmbedEP} to represent vulnerability-related posts from underground forums, encode the word features into a fixed-size vector space. In line with prior work, we use the Doc2Vec model~\cite{le2014distributed} to learn the embeddings on the document from the training set.
We use separate models on the NVD descriptions, Write-ups, PoC Info and the comments from the PoC Code artifacts. 
We perform grid search for the hyper-parameters of the model, and report the performance of the best-performing models.
The 200-dimensional vectors are obtained from the distributed bag of words (D-BOW) algorithm trained over 50 epochs, using a windows size of 4, a sampling threshold of 0.001, using the sum of the context words, and a frequency threshold of 2.

In Figure~\ref{fig:perf_f} we compare the effect of the alternative NLP features on \texttt{EE}.
First, we observe that TF-IDF does not improve the performance over unigrams.
This suggests that our classifier does not require term frequency to learn the vulnerability characteristics reflected through artifacts, which seems to even hurt performance slightly.
This can be explained intuitively, as different artifacts frequently reuse the same jargon for the same vulnerability, but the number of distinct artifacts is not necessarily correlated with exploitability.
However, the TF-IDF classifier might over-emphasize the numerical value of these features, rather than learning their presence. 

Surprisingly, the embedding features result in a significant performance drop, in spite of our hyper-parameter tuning attempts.
We observe that the various natural language artifacts in our corpus are long and verbose, resulting in a large number of tokens that need to be aggregated into a single embedding vector.
Due to this aggregation and feature compression, the distinguishing words which indicate exploitability might not remain sufficiently expressive within the final embedding vector that our classifier uses as inputs.
While our results do not align with the DarkEmbed work finding that paragraph embeddings outperform simpler features, we note that DarkEmbed is primarily using posts from underground forums, which the authors report are shorter than public write-ups.
Overall, our result reveals that creating higher level, semantic, NLP features for exploit prediction is a challenging problem, and requires solutions beyond using off-the-shelf tools. 
We leave this problem to future work.

\begin{figure}[t]
\centering
\includegraphics[height=2.1in]{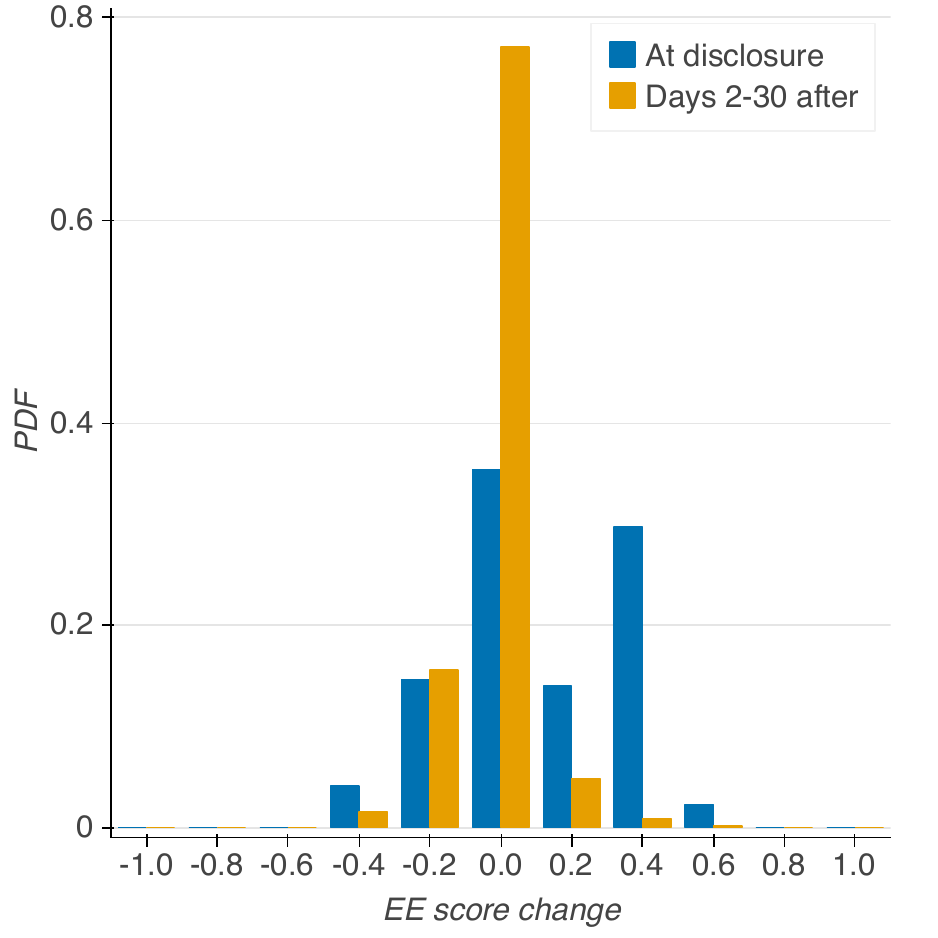}
\caption{The distribution of EE score changes, at the time of disclosure and on all events within 30 days after disclosure.}
\label{fig:ee_change_distribution}
\end{figure}

\topic{\texttt{EE} is stable over time} 
In order to observe how \texttt{EE} is influenced by the publication of various artifacts, we look at the changes in the score of the classifier.
Figure~\ref{fig:ee_change_distribution} plots, for the entire test set, the distribution of score changes in two cases: at the time of disclosure compared to an instance with no features, and from the second to the $30^{th}$ day after disclosure, on days where there were artifacts published for an instance.
We observe that at the time of disclosure, the classifier changes drastically, shifting the instance towards either 0.0 or 1.0, while the large magnitude of the shifts indicate a high confidence.
However, we observe that artifacts published on subsequent days have a much different effect.
In 79\% of cases, published artifacts have almost no effect on changing the classification score, while the remaining 21\% of events are the primary drivers of score changes.
The two observations allow us to conclude that artifacts published at the time of disclosure contain some of the most informative features, and that \texttt{EE} is stable over time, its evolution being determined by few consequential artifacts.

\begin{figure}[t]
    \begin{subfigure}{.24\textwidth}
        \centering
        \includegraphics[height=1.8in]{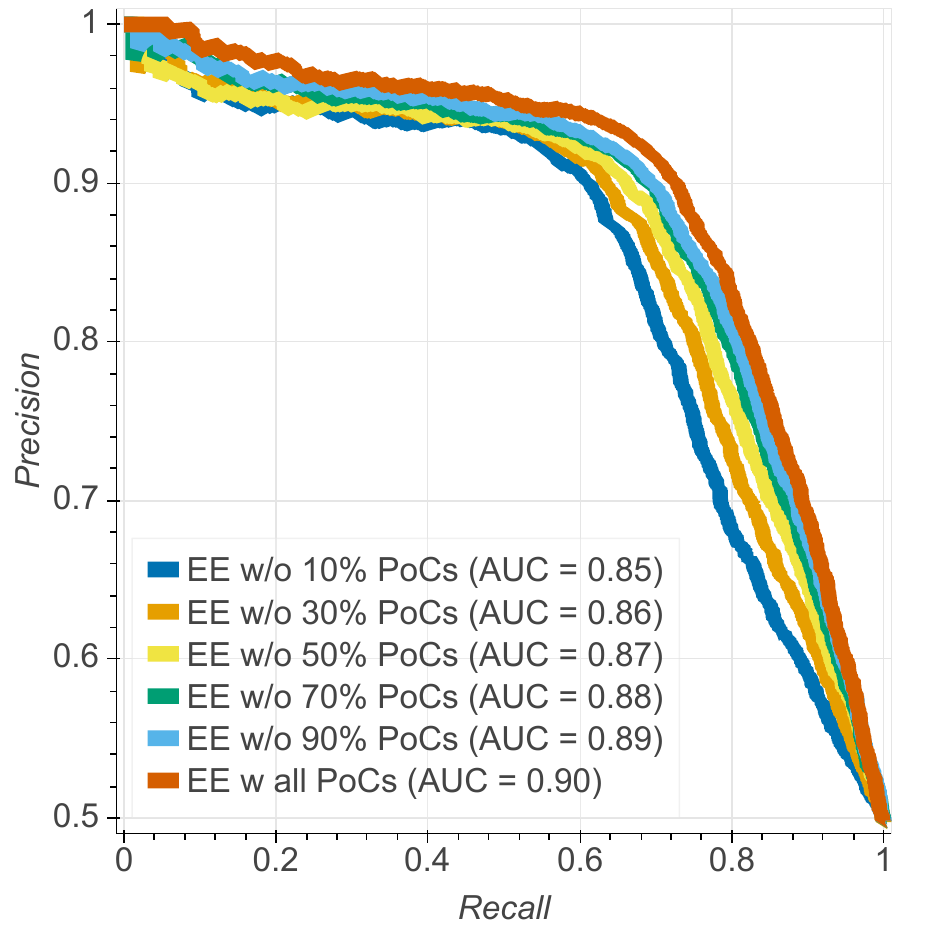}
        \vspace{-0.2in}
        \caption{}
        \label{fig:missing_pocs_pr}
    \end{subfigure}%
    \begin{subfigure}{.24\textwidth}
        \centering
        \includegraphics[height=1.8in]{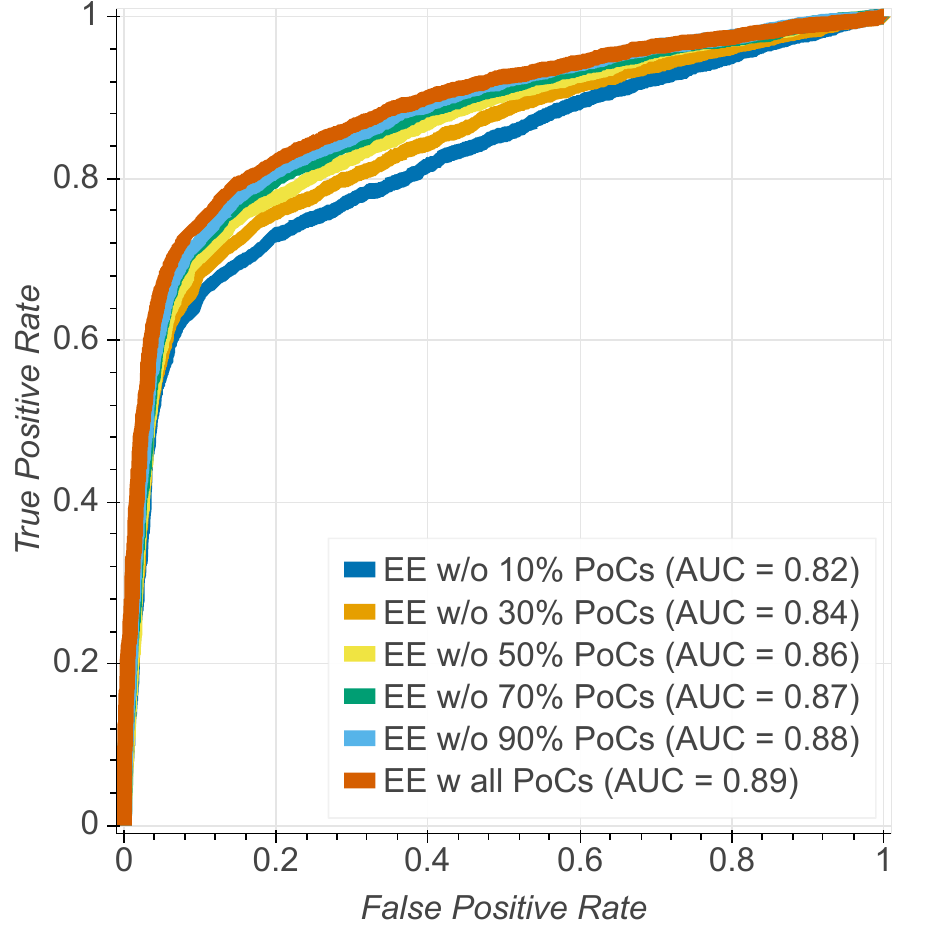}
        \vspace{-0.2in}
        \caption{}
        \label{fig:missing_pocs_auc}
    \end{subfigure}%
    \caption{Performance of our classifier when a fraction of the PoCs is missing.}
    \label{fig:missing_pocs}
\end{figure}

\topic{\texttt{EE} is robust to missing exploits} 
In order to observe how \texttt{EE} performs when some of the PoCs are missing, we simulate a scenario in which a varying fraction of them are not seen at test-time for vulnerabilities in \texttt{DS1}.
The results are plotted in Figure~\ref{fig:missing_pocs}, and highlight that, even if a significant fraction of PoCs is missing, our classifier is able to utilize the other types of artifacts to maintain a high performance.   

}

\begin{table}[t]
\renewcommand{\arraystretch}{1.25}
\setlength{\tabcolsep}{4.2pt}
\centering
\resizebox{1.0\columnwidth}{!}{
\scriptsize
\begin{tabular}{ |c | c | c | c | c | c | c |}
  \hline
 & \tiny$\mathcal{P}^{CVSS}$ & \tiny$\mathcal{P}^{EPSS}$ & \tiny$\mathcal{P}^{EE}(\delta)$ &  \tiny$\mathcal{P}^{EE}(\delta+\mbox{10})$ & \tiny$\mathcal{P}^{EE}(\delta+\mbox{30})$ &  \tiny$\mathcal{P}^{EE}(\mbox{2020-12-07})$ \\ \hline
 \tiny\textbf{Mean} & 0.51 & 0.36 & 0.31 & 0.25 & 0.22 & 0.04 \\ \hline
 \tiny\strut\textbf{Std} &  0.24 & 0.28 & 0.33 & 0.25 & 0.27 & 0.11  \\ \hline
 \tiny\strut\textbf{Median} & 0.35 & 0.40 & 0.22 & 0.12 & 0.10 & 0.00  \\ \hline
\end{tabular}}
\caption{Performance of \texttt{EE} and baselines at prioritizing critical vulnerabilities. $\mathcal{P}$ captures the fraction of recent non-exploited vulnerabilities scored higher than critical ones.}
\label{table:important_cveids_results}
\end{table}

\subsection{Case Studies}
\label{sec:casestudy}
In this section we investigate the practical utility of \texttt{EE} through \usenix{two}\arxiv{three} case studies.

\topic{\texttt{EE} for critical vulnerabilities}
To understand how well \texttt{EE} distinguishes important vulnerabilities, we measure its performance on a list of recent ones flagged for prioritized remediation by FireEye~\cite{fireeyelist}. 
The list was published on December 8, 2020, after the corresponding functional exploits were stolen~\cite{fireeyelistblog}. 
Our dataset contains 15 of the 16 critical vulnerabilities. 

We measure how well our classifier prioritizes these vulnerabilities compared to static baselines, using the $\mathcal{P}$ prioritization metric defined in the previous section, which computes the fraction of non-exploited vulnerabilities from a set $S$
that are scored higher than the critical ones.
For each of the 15 vulnerabilities, we set $S$ to contain all others disclosed within 30 days from it, which represent the most frequent alternatives for prioritization decisions.  
Table~\ref{table:important_cveids_results} compares the statistics for the baselines, and for $\mathcal{P}^{EE}$ computed on the date critical vulnerabilities were disclosed $\delta$, 10 and 30 days later, as well as one day before the prioritization recommendation was published.
CVSS scores are published a median of 18 days after disclosure, and we observe that \texttt{EE} already outperforms static baselines based only on the features available at disclosure, while time-varying features improve performance significantly.
Overall, one day before the prioritization recommendation is issued, our classifier scores the critical vulnerabilities below only 4\% of these with no known exploit.
\usenix{
In our technical report~\cite{suciu2021expected} we list the individual vulnerabilities, their scores, and analyze the factors impacting performance for various examples.
}
\arxiv{
Table~\ref{table:important_cveids_results_samecpe_apx} shows the performance statistics of our classifier when $|S|$ contains only vulnerabilities published within 30 days of the critical ones and that affect the same products as the critical ones.
The result further highlights the utility of \texttt{EE}, as its ranking outperforms baselines and prioritizes the most critical vulnerabilities for a particular product.

\begin{table}[t]
\renewcommand{\arraystretch}{1.25}
\setlength{\tabcolsep}{4.2pt}
\centering
\resizebox{1.0\columnwidth}{!}{
\scriptsize
\begin{tabular}{ |c | c | c | c | c | c | c |}
  \hline
 & \tiny$\mathcal{P}^{CVSS}$ & \tiny$\mathcal{P}^{EPSS}$ & \tiny$\mathcal{P}^{EE}(\delta)$ &  \tiny$\mathcal{P}^{EE}(\delta+\mbox{10})$ & \tiny$\mathcal{P}^{EE}(\delta+\mbox{30})$ &  \tiny$\mathcal{P}^{EE}(\mbox{2020-12-07})$ \\ \hline
 \tiny\textbf{Mean} & 0.51 & 0.42 & 0.34 & 0.34 & 0.23 & 0.11 \\ \hline
 \tiny\strut\textbf{Std} &  0.39 & 0.32 & 0.40 & 0.40 & 0.30 & 0.26  \\ \hline
 \tiny\strut\textbf{Median} & 0.43 & 0.35 & 0.00 & 0.04 & 0.14 & 0.00  \\ \hline
\end{tabular}}
\caption{Performance of \texttt{EE} and baselines at prioritizing critical vulnerabilities. $\mathcal{P}$ captures the fraction of recent non-exploited vulnerabilities for the same products and scored higher than critical ones.}
\label{table:important_cveids_results_samecpe_apx}
\end{table}

Table~\ref{table:important_cveids_apx} lists the 15 out of 16 critical vulnerabilities in our dataset flagged by FireEye. The table lists the estimated disclosure date, the number of days after disclosure when CVSS was published, and when exploitation evidence emerged.
Table~\ref{table:important_cveids_results_apx} contains the per-vulnerability performance of our classifier for all 15 vulnerabilities when $|S|$ contains vulnerabilities published within 30 days of the critical ones.
Below, we manually analyze some of the 15 vulnerabilities in more details by combining \texttt{EE} and $\mathcal{P}$.

\textbf{CVE-2019-0604:}
Table~\ref{table:important_cveids_results_apx} shows the performance of our classifier on CVE-2019-0604, which improves when more information becomes publicly available. 
At the disclosure time, there is only one available write-up which yields a low \texttt{EE} because it contains little descriptive features. 
23 days later, when NVD descriptions become available, \texttt{EE} decreases even further. 
However, two technical write-ups on days 87 and 352 result in sharp increases of \texttt{EE}, from 0.03 to 0.22 and to 0.78 respectively. 
This is because they contain detailed technical analyses of the vulnerability, which our classifier interprets as an increased exploitation likelihood. 

\textbf{CVE-2019-8394:}
$\mathcal{P}$ fluctuates between 0.82 and 0.24 on CVE-2019-8394. 
At disclosure time, this vulnerability gathers only one write-up, and our classifier outputs a low \texttt{EE}. 
From disclosure time to day 10, there are two small changes in \texttt{EE}, but at day 10, when NVD information is available, there is a sharp decrease on \texttt{EE} from 0.12 to 0.04. From day 10 to day 365, \texttt{EE} does not change anymore due to no more information added. 
The decrease of \texttt{EE} at day 10 explains the sharp jump between $\mathcal{P}^{EE}(\mbox{0})$ and $\mathcal{P}^{EE}(\mbox{10})$ but not the fluctuations after $\mathcal{P}^{EE}(\mbox{10})$. 
This is caused by the \texttt{EE} of other vulnerabilities disclosed around the same period, which our classifier ranks higher than CVE-2019-8394.

\textbf{CVE-2020-10189 and CVE-2019-0708:}
These two vulnerabilities receive high \texttt{EE} throughout the entire observation period, due to detailed technical information available at disclosure, which allows our classifier to make confident predictions. 
CVE-2019-0708 gathers 35 write-ups in total, and 4 of them are available at disclosure. 
Though CVE-2020-10189 only gathers 4 write-ups in total, 3 of them are available within 1 day of disclosure and contained informative features. 
These two examples show that our classifier benefits from an abundance of informative features published early on, and this information contribute to confident predictions that remain stable over time.

\begin{table}[t]
\renewcommand{\arraystretch}{1.25}
\centering
\begin{tabular}{ | c | c | c | c |}
  \hline
 \small\textbf{CVE-ID} & \small\strut\textbf{Disclosure}  & \small\strut\textbf{CVSS Delay} & \small\strut\textbf{Exploit Delay}\\ \hline
2019-11510 & 2019-04-24 & 15 & 125 \\ \hline
2018-13379 & 2019-03-24 & 73 & 146 \\ \hline
2018-15961 & 2018-09-11 & 66 & 93 \\ \hline
2019-0604 & 2019-02-12 & 23 & 86 \\ \hline
2019-0708 & 2019-05-14 & 2 & 8 \\ \hline
2019-11580 & 2019-05-06 & 28 & ? \\ \hline
2019-19781 & 2019-12-13 & 18 & 29 \\ \hline
2020-10189 & 2020-03-05 & 1 & 5 \\ \hline
2014-1812 & 2014-05-13 & 1 & 1 \\ \hline
2019-3398 & 2019-03-31 & 22 & 19 \\ \hline
2020-0688 & 2020-02-11 & 2 & 16 \\ \hline
2016-0167 & 2016-04-12 & 2 & ? \\ \hline
2017-11774 & 2017-10-10 & 24 & ? \\ \hline
2018-8581 & 2018-11-13 & 34 & ? \\ \hline
2019-8394 & 2019-02-12 & 10 & 412 \\ \hline
\end{tabular}
\caption{List of exploited CVE-IDs in our dataset recently flagged for prioritized remediation. Vulnerabilities where exploit dates are unknown are marked with '?'.}
\label{table:important_cveids_apx}
\end{table}

\begin{table*}[t]
\renewcommand{\arraystretch}{1.25}
\centering
\begin{tabular}{ | c || c | c || c | c | c | c | }
  \hline
 \small\textbf{CVE-ID} & \small\strut\textbf{$\mathcal{P}^{CVSS}$}  & \small\strut\textbf{$\mathcal{P}^{EPSS}$} & \small\strut\textbf{$\mathcal{P}^{EE}(\mbox{0})$ } & \small\strut\textbf{$\mathcal{P}^{EE}(\mbox{10})$ } & \small\strut\textbf{$\mathcal{P}^{EE}(\mbox{30})$ } & \small\strut\textbf{$\mathcal{P}^{EE}(\mbox{2020-12-07})$}\\ \hline
2014-1812 & 0.81 & 0.48 & 0.00 & 0.01 & 0.03 & 0.03 \\ \hline
2016-0167 & 0.97 & 0.15 & 0.79 & 0.50 & 0.13 & 0.04 \\ \hline
2017-11774 & 0.61 & 0.12 & 0.99 & 0.13 & 0.23 & 0.08 \\ \hline
2018-13379 & 0.28 & 0.42 & 0.00 & 0.06 & 0.06 & 0.00 \\ \hline
2018-15961 & 0.25 & 0.55 & 0.39 & 0.46 & 0.41 & 0.00 \\ \hline
2018-8581 & 0.64 & 0.30 & 0.42 & 0.29 & 0.26 & 0.01 \\ \hline
2019-0604 & 0.34 & 0.54 & 0.73 & 0.62 & 0.80 & 0.01 \\ \hline
2019-0708 & 0.30 & 0.07 & 0.00 & 0.00 & 0.00 & 0.00 \\ \hline
2019-11510 & 0.34 & 0.85 & 0.45 & 0.41 & 0.61 & 0.00 \\ \hline
2019-11580 & 0.32 & 0.89 & 0.04 & 0.06 & 0.01 & 0.02 \\ \hline
2019-19781 & 0.36 & 0.01 & 0.09 & 0.13 & 0.00 & 0.00 \\ \hline
2019-3398 & 0.82 & 0.40 & 0.67 & 0.30 & 0.10 & 0.00 \\ \hline
2019-8394 & 0.69 & 0.07 & 0.22 & 0.82 & 0.76 & 0.48 \\ \hline
2020-0688 & 0.77 & 0.62 & 0.00 & 0.00 & 0.00 & 0.00 \\ \hline
2020-10189 & 0.24 & 0.01 & 0.00 & 0.00 & 0.00 & 0.00 \\ \hline
\end{tabular}
\caption{The performance of baselines and \texttt{EE} at prioritizing critical vulnerabilities.}
\label{table:important_cveids_results_apx}
\end{table*}
}
Our results indicate that \texttt{EE} is a valuable input to patching prioritization frameworks, because it outperforms existing metrics and improves over time.

\begin{figure}[t]
    \begin{subfigure}{.24\textwidth}
        \centering
        \includegraphics[height=1.7in]{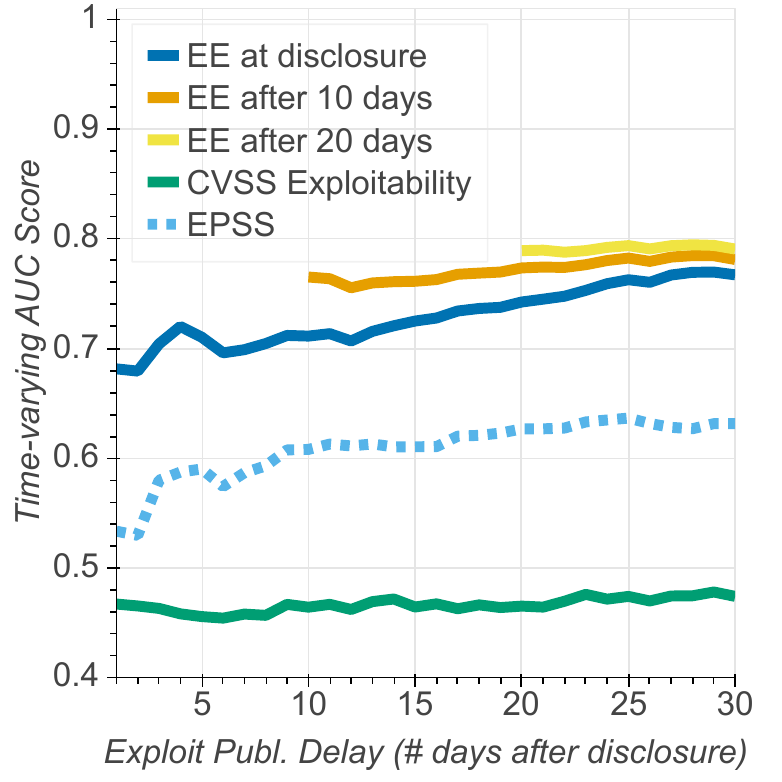}
        \vspace{-0.2in}
        \caption{}
        \label{fig:auc_score_a}
    \end{subfigure}%
    \begin{subfigure}{.24\textwidth}
        \centering
        \includegraphics[height=1.7in]{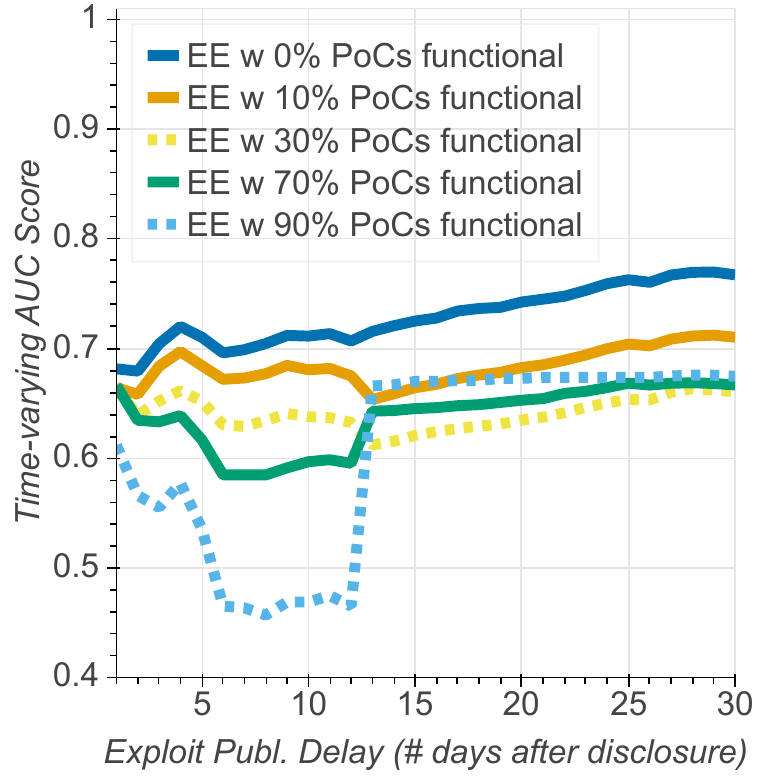}
        \vspace{-0.2in}
        \caption{}
        \label{fig:auc_score_b}
    \end{subfigure}%
    \caption{Time-varying AUC when distinguishing exploits published within $t$ days from disclosure (a) for \texttt{EE} and baselines, (b) simulating earlier exploit availability.}
    \label{fig:auc_score}
\end{figure}

\topic{\texttt{EE} for emergency response} 
Next, we investigate how well our classifier can predict exploits published shortly after disclosure.
To this end, we look at the 924 vulnerabilities in \texttt{DS3} for which we obtained exploit publication estimates. 
\usenix{
In the technical report~\cite{suciu2021expected} we perform a statistical test and conclude that \texttt{DS3} is a representative sample of all other exploits in our dataset. 
}
\arxiv{
To test whether the vulnerabilities in \texttt{DS3} are a representative sample of all other exploits, we perform a two-sample test, under the null hypothesis that vulnerabilities in \texttt{DS3} and exploited vulnerabilities in \texttt{DS2} which are not in \texttt{DS3} are drawn from the same distribution.
Because instances are multivariate and our classifier learns feature representations for these vulnerabilities, we use a technique designed for this scenario, called Classifier Two-Sample Tests(C2ST)~\cite{lopez2016revisiting}.
C2ST repeatedly trains classifiers to distinguish between instances in the two samples and, using a Kolmogorov-Smirnoff test, compares the probabilities assigned to instances from the two, in order to determine whether any statistically significant difference can be established between them.
We apply C2ST on the features learned by our classifier (the last hidden layer which contains 100 dimensions), and we were unable to reject the null hypothesis that the two samples are drawn from the same distribution, at p=0.01.
Based on this result, we conclude that \texttt{DS3} is a representative sample of all other exploits in our dataset. 
This means that, when considering the features evaluated in our paper, we find no evidence of biases in \texttt{DS3}.
}

%

We measure the performance of \texttt{EE} at predicting vulnerabilities exploited within $t$ days from disclosure.
For a given vulnerability $i$ and $EE_i(z)$ computed on date $z$, we can compute the time-varying sensitivity $Se=P(EE_{i}(z)>c|D_{i}(t)=1)$ and specificity $Sp=P(EE_{i}(z)\leq c|D_{i}(t)=0)$~\cite{heagerty2000time}, where $D_{i}(t)$ indicates whether the vulnerability was already exploited by time $t$.
By varying the detection threshold $c$, we obtain the time-varying AUC of the classifier which reflects how well the classifier separates exploits happening within $t$ days from these happening later on.
In Figure~\ref{fig:auc_score_a} we plot the AUC for our classifier evaluated on the day of disclosure $\delta$, as well as 10 and 20 days later, for the exploits published within 30 days. 
While the CVSS Exploitability remains below 0.5, $EE(\delta)$ constantly achieves an AUC above 0.68.
This suggests that the classifiers implicitly learns to assign higher scores to vulnerabilities that are exploited sooner than to these exploited later.
%
For $EE(\delta+10)$ and $EE(\delta+20)$, in addition to similar trends over time, we also observe the benefits of additional features collected in the days after disclosure, which shift the overall prediction performance upward.

We further consider the possibility that the timestamps in \texttt{DS3} may be affected by label noise. 
We evaluate the potential impact of this noise with an approach similar to the one in Section~\ref{sec:feature-dependent-noise-remediation}.
%
%
We simulate scenarios where we assume that a percentage of PoCs are already functional, which means that their later exploit-availability dates in \texttt{DS3} are incorrect.
For those vulnerabilities, we update the exploit availability date to reflect the publication date of these PoCs.
This provides a conservative estimate, because the mislabeled PoCs could be in an advanced stage of development, but not yet fully functional, and the exploit-availability dates could also be set too early.
We simulate percentages of late timestamps ranging from 10--90\%.
Figure~\ref{fig:auc_score_b} plots the performance of $EE(\delta)$ in this scenario, averaged over 5 repetitions.
We observe that even if 70\% of PoCs are considered functional, the classifier outperforms the baselines and maintains an AUC above 0.58, 
Interestingly, performance drops after disclosure and is affected the most on predicting exploits published within 12 days.
Therefore, the classifier based on disclosure-time artifacts learns features of easily exploitable vulnerabilities, which get published immediately, but does not fully capture the risk of functional PoC that are published early. 
We mitigate this effect by updating \texttt{EE} with new artifacts daily, after disclosure. 
Overall, the result suggests that $EE$ may be useful in emergency response scenarios, where it is critical to urgently patch the vulnerabilities that are about to receive functional exploits. 

\arxiv {

\begin{table}[t]
\renewcommand{\arraystretch}{1.25}
\resizebox{\columnwidth}{!}{
\begin{tabular}{ | c | c | c | }
  \hline
 & \textbf{Player 1} & \textbf{Player 2} \\ \hline
Loss if attacked  & $l_1(t) = 5000, \forall t$ & $l_2(t) = 500, \forall t$ \\ \hline
Patching rate & $h_1(t) = 1 - 0.8^t, \forall t$ & $h_2(t) = 1-0.8^t, \forall t$ \\ \hline
\end{tabular}%
}
\caption{Cyber-warfare game simulation parameters.}
\label{table:gametheory_setup}
\end{table}

\topic{\texttt{EE} for vulnerability mitigation} To investigate the practical utility of \texttt{EE}, we conduct a case study of vulnerability mitigation strategies. One example of vulnerability mitigation is cyber warfare, where nations acquire exploits and make decisions based on new vulnerabilities. 
Existing cyber-warfare research such as that by Bao et al.\cite{bao2017shall} rely on knowledge of exploitability for game strategies.
For these models, it is therefore crucial that exploitability estimates are \emph{timely and accurate}, because inaccuracies could lead to sub-optimal strategies.  
Because these requirements match our design decisions for learning \texttt{EE}, we evaluate its effectiveness  in the context of a cyber-game. 
We simulate the case of CVE-2017-0144, the vulnerability targeted by the EternalBlue exploit~\cite{eternalblue}.
The game has two players, where Player 1, a government, possesses an exploit that gets stolen, and Player 2, an evil hacker who might know about it could purchase it or re-create it.
We set the game parameters aligned with the real-world circumstances for the EternalBlue vulnerability, shown in Table~\ref{table:gametheory_setup}.
In this setup, Player 1's loss of being attacked is significantly greater than Player 2's, because a government needs to take into account the loss for a large population, as opposed to that for a small group or an individual.
Both players begin patching once the vulnerability is disclosed, at round 0.
The patching rates, which are the cumulative proportion of vulnerable resources being patched over time, are equal for both players and follow the pattern measured in prior work~\cite{Bilge12:ZeroDay}.
We also assume that the exploit becomes available at $t=31$, as this corresponds to the delay after which EternalBlue was published.

Our experiment assumes that Player 1 uses the cyber-warfare model~\cite{bao2017shall} to compute whether they should attack Player 2 after vulnerability disclosure.
The calculation requires Player 2's exploitability, which we assign using two approaches: The CVSS Exploitability score normalized to 1 (which yields a value of 0.55), and the time-varying \texttt{EE}.
Our classifier outputs an exploitability of 0.94 on the day of disclosure, and updates it to 0.97 three days later, only to maintain it constant afterwards.
We compute the optimal strategy for the two approaches, and we compare them using the resulting utility for Player 1.

Figure~\ref{fig:gametheory} shows that the strategy associated with \texttt{EE} is preferable over the CVSS one.
Although Player 1 will inevitably lose in the game  (because they have a much larger vulnerable population), \texttt{EE} improves Player 1's utility by 10\%.
Interestingly, we find that \texttt{EE} also changes Player 1's strategy to towards a more aggressive one.
This is because \texttt{EE} is updated when more information emerges, which in turn increases the expected exploitability assumed for Player 2.
When Player 2 is unlikely to have a working exploit, Player 1 would not attack because that may leak information on how to weaponize the vulnerability, and Player 2 may convert the received exploit to an inverse attack~\cite{bao2017your}.
As Player 2's exploitability increases, Player 1 will switch to attacking because it is likely that Player 2 already possesses an exploit.
Therefore, an increasing exploitability pushes Player 1 towards a more aggressive strategy.

\begin{figure}[t]
\centering
\includegraphics[height=1.8in]{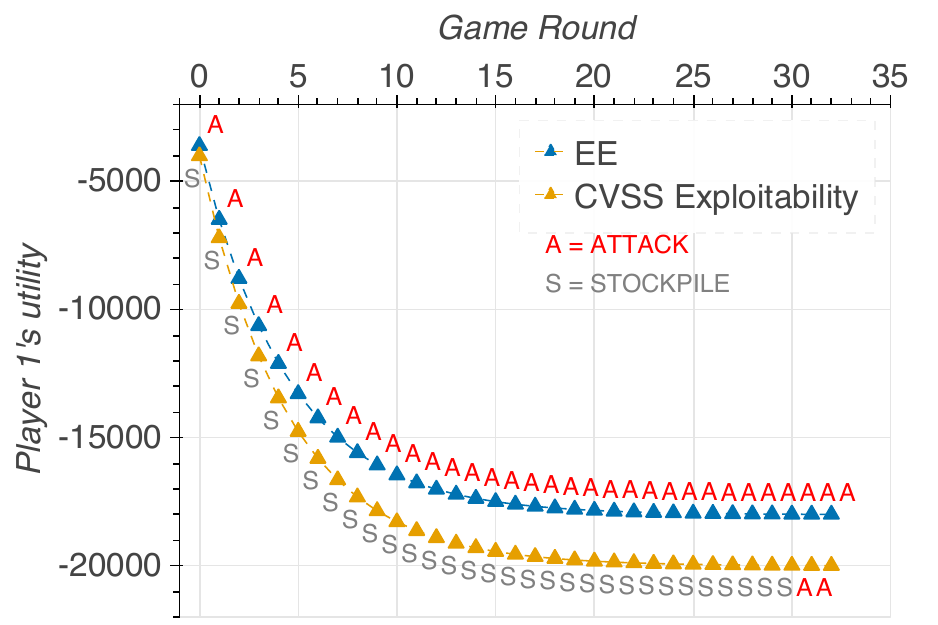}
\caption{Result of our cyber-warfare game simulation. The utility of the player is improved by 2,000 points when using \texttt{EE}. The player actions are also different, the CVSS player only attacking when the exploit is leaked, in round 31.}
\label{fig:gametheory}
\end{figure}

}

\section{Related Work}
\label{sec:related-work}

\topic{Predicting exploits in the wild} 
Most of the prior exploit prediction work has been towards the tangential task of predicting exploits in the wild. This has been investigated in our prior study~\cite{USENIX-Security-2015} and Chen et al.~\cite{chen2019using} by monitoring Twitter for vulnerability discussions, and Xiao et al.~\cite{xiao2018patching} by using post-disclosure field data about exploitation. Jacobs et al.~\cite{WEIS-19-Exploits} used vulnerability prevalence data to improve prediction. Jacobs et al.~\cite{jacobs2019exploit} proposed EPSS, a scoring system for exploits. Allodi~\cite{allodi2017economic} calculated the likelihood of observing exploits in the wild after they are traded in underground forums.


\topic{Vulnerability Exploitability} 
Allodi and Massacci~\cite{Allodi12:VulnerabilityScores} investigated the utility of the CVSS scores for capturing the likelihood of attacks in the wild.
Prior work by Bozorgi et al.~\cite{DBLP:conf/kdd/BozorgiSSV10} formulated exploitability estimation as the problem of predicting the existence of PoCs based on vulnerability characteristics. 
Allodi and Massaci~\cite{allodi2014comparing} concluded that the publication of a PoC in ExploitDB is not a good indicator for exploits in the wild.
Our work shows that, while their presence might not be a sufficiently accurate indicator, the features within these PoCs are useful for predicting functional exploits.
DarkEmbed~\cite{Tavabi2018DarkEmbedEP} uses natural language models trained on private data from underground forum discussions to predict the availability of exploits, but such artifacts are generally published with a delay~\cite{allodi2017economic}. Instead, \texttt{EE} uses only publicly available artifacts for predicting exploits soon after disclosure; we were unable to obtain these artifacts for comparison upon contacting the authors.

\topic{PoCs} PoCs were also investigated in measurements on vulnerability lifecycles.  Shahzad et al~\cite{shahzad:12} performed a measurement of the vulnerability lifecycle, discovering that PoCs are generally released at the time of disclosure.
Mu. et al~\cite{mu2018understanding} manually curated and utilized PoCs to trigger the vulnerabilities.
FUZE~\cite{wu2018fuze} used PoCs to aid exploit generation for 5 functional exploits.

\topic{Label Noise}
The problem of label noise has been studied extensively in machine learning~\cite{frenay2013classification}, primarily focusing on random and class-dependent noise. Limited work focuses on feature-dependent noise, which requires either strong theoretical guarantees about the sample space~\cite{menon2016learning} or depends on a large dataset of clean labels to learn noise probabilities~\cite{xiao2015learning}.
In security, the closest to our work is the study by Deloach et al~\cite{deloach2016android} which models noise in malware classification as class-dependent.

%
%

\section{Conclusions}
\label{sec:conclusions}
By investigating exploitability as a time-varying process, we discover that it can be learned using supervised classification techniques and updated continuously.
We discover three challenges associated with exploitability prediction.
First, it is prone to feature-dependent label noise, a type considered by the machine learning community as the most challenging.
Second, it needs new categories of features, as it differs qualitatively from the related task of predicting exploits in the wild.
Third, it requires new metrics for performance evaluation, designed to capture practical vulnerability prioritization considerations.  

We design the \texttt{EE} metric, which, on a dataset of 103,137 vulnerabilities, improves precision from 49\% to 86\% over state-of-the art predictors.
\texttt{EE} can learn to mitigate feature-dependent label noise, capitalizes on highly predictive features that we extract from PoCs and write-ups, improves over time, and has practical utility in predicting imminent exploits and prioritizing critical vulnerabilities.
\topic{Acknowledgments}
We thank Vulners and Frank Li for their data. We also thank the anonymous reviewers, Yigitcan Kaya and Ben Edwards for feedback. 
This material was supported by a grant from the Department of Defense, the Army Research Office (W911NF-17-1-0370), the National Science Foundation (CNS-2000792), and based upon work supported by the Defense Advanced Research Projects Agency (DARPA) under Agreement No. HR00112190093. Approved for public release; distribution is unlimited.

{



\bibliographystyle{abbrv}


\bibliography{bibliography/security}

\begin{thebibliography}{10}

\bibitem{allodi2017economic}
L.~Allodi.
\newblock Economic factors of vulnerability trade and exploitation.
\newblock In {\em Proceedings of the 2017 ACM SIGSAC Conference on Computer and
  Communications Security}, pages 1483--1499. ACM, 2017.

\bibitem{Allodi12:VulnerabilityScores}
L.~Allodi and F.~Massacci.
\newblock A preliminary analysis of vulnerability scores for attacks in wild.
\newblock In {\em CCS BADGERS Workshop}, Raleigh, NC, Oct 2012.

\bibitem{allodi2014comparing}
L.~Allodi and F.~Massacci.
\newblock Comparing vulnerability severity and exploits using case-control
  studies.
\newblock {\em ACM Transactions on Information and System Security (TISSEC)},
  17(1):1, 2014.

\bibitem{attack_signatures}
{Symantec attack signatures}.
\newblock \url{https://www.symantec.com/security_response/attacksignatures/}.

\bibitem{bao2017shall}
T.~Bao, Y.~Shoshitaishvili, R.~Wang, C.~Kruegel, G.~Vigna, and D.~Brumley.
\newblock How shall we play a game?: a game-theoretical model for cyber-warfare
  games.
\newblock In {\em 2017 IEEE 30th Computer Security Foundations Symposium
  (CSF)}, pages 7--21. IEEE, 2017.

\bibitem{bao2017your}
T.~Bao, R.~Wang, Y.~Shoshitaishvili, and D.~Brumley.
\newblock Your exploit is mine: Automatic shellcode transplant for remote
  exploits.
\newblock In {\em 2017 IEEE Symposium on Security and Privacy (SP)}, pages
  824--839. IEEE, 2017.

\bibitem{Bilge12:ZeroDay}
L.~Bilge and T.~Dumitra\cb{s}.
\newblock Before we knew it: an empirical study of zero-day attacks in the real
  world.
\newblock In {\em ACM Conference on Computer and Communications Security},
  pages 833--844, 2012.

\bibitem{bleepingcomp-ekits}
{BleepingComputer}.
\newblock Ie zero-day adopted by rig exploit kit after publication of poc code.
\newblock
  \url{https://www.bleepingcomputer.com/news/security/ie-zero-day-adopted-by-rig-exploit-kit-after-publication-of-poc-code/},
  2018.

\bibitem{DBLP:conf/kdd/BozorgiSSV10}
M.~Bozorgi, L.~K. Saul, S.~Savage, and G.~M. Voelker.
\newblock Beyond heuristics: learning to classify vulnerabilities and predict
  exploits.
\newblock In {\em KDD}, Washington, DC, Jul 2010.

\bibitem{securityfocus}
{Bugtraq}.
\newblock Securityfocus.
\newblock \url{https://www.securityfocus.com/}, 2019.

\bibitem{caliskan2015anonymizing}
A.~Caliskan-Islam, R.~Harang, A.~Liu, A.~Narayanan, C.~Voss, F.~Yamaguchi, and
  R.~Greenstadt.
\newblock De-anonymizing programmers via code stylometry.
\newblock In {\em 24th USENIX Security Symposium (USENIX Security 15)}, pages
  255--270, 2015.

\bibitem{chakraborty2018adversarial}
A.~Chakraborty, M.~Alam, V.~Dey, A.~Chattopadhyay, and D.~Mukhopadhyay.
\newblock Adversarial attacks and defences: A survey.
\newblock {\em arXiv preprint arXiv:1810.00069}, 2018.

\bibitem{chen2019using}
H.~Chen, R.~Liu, N.~Park, and V.~Subrahmanian.
\newblock Using twitter to predict when vulnerabilities will be exploited.
\newblock In {\em Proceedings of the 25th ACM SIGKDD International Conference
  on Knowledge Discovery \& Data Mining}, pages 3143--3152, 2019.

\bibitem{MITRE:CWE}
T.~M. Corporation.
\newblock Common weaknesses enumeration.
\newblock \url{https://cwe.mitre.org}.

\bibitem{crowdstrike_notpetya}
NotPetya Technical Analysis.
\newblock
  \url{https://www.crowdstrike.com/blog/petrwrap-ransomware-technical-analysis-triple-threat-file-encryption-mft-encryption-credential-theft/}.

\bibitem{d2pack}
{D2 Security}.
\newblock D2 exploitation pack.
\newblock \url{https://www.d2sec.com/pack.html}, 2019.

\bibitem{deloach2016android}
J.~DeLoach, D.~Caragea, and X.~Ou.
\newblock Android malware detection with weak ground truth data.
\newblock In {\em 2016 IEEE International Conference on Big Data (Big Data)},
  pages 3457--3464. IEEE, 2016.

\bibitem{dullien2017weird}
T.~F. Dullien.
\newblock Weird machines, exploitability, and provable unexploitability.
\newblock {\em IEEE Transactions on Emerging Topics in Computing}, 2017.

\bibitem{dunn1961multiple}
O.~J. Dunn.
\newblock Multiple comparisons among means.
\newblock {\em Journal of the American statistical association},
  56(293):52--64, 1961.

\bibitem{eiram2013exploitability}
C.~Eiram.
\newblock Exploitability/priority index rating systems (approaches, value, and
  limitations), 2013.

\bibitem{MS:ExploitabilityIndex}
Microsoft exploitability index.
\newblock Microsoft, 30 March 2020.
\newblock \url{https://www.microsoft.com/en-us/msrc/exploitability-index}.

\bibitem{exploitdb}
{ExploitDB}.
\newblock The exploit database.
\newblock \url{https://www.exploit-db.com/}, 2019.

\bibitem{fireeyelist}
FireEye.
\newblock Fireeye red team tools.
\newblock
  \url{https://github.com/fireeye/red_team_tool_countermeasures/blob/master/CVEs_red_team_tools.md}.

\bibitem{fireeyelistblog}
FireEye.
\newblock Unauthorized access of fireeye red team tools.
\newblock
  \url{https://www.fireeye.com/blog/threat-research/2020/12/unauthorized-access-of-fireeye-red-team-tools.html}.

\bibitem{frenay2013classification}
B.~Fr{\'e}nay and M.~Verleysen.
\newblock Classification in the presence of label noise: a survey.
\newblock {\em IEEE transactions on neural networks and learning systems},
  25(5):845--869, 2013.

\bibitem{githublinguist}
{GitHub}.
\newblock linguist.
\newblock \url{https://github.com/github/linguist}, 2020.

\bibitem{hackerone:disclosure}
Hackerone - disclosure guidelines.
\newblock HackerOne, 30 March 2009.
\newblock \url{https://www.hackerone.com/disclosure-guidelines}.

\bibitem{heagerty2000time}
P.~J. Heagerty, T.~Lumley, and M.~S. Pepe.
\newblock Time-dependent roc curves for censored survival data and a diagnostic
  marker.
\newblock {\em Biometrics}, 56(2):337--344, 2000.

\bibitem{VEPdocument}
T.~W. House.
\newblock Vulnerabilities equities policy and process for the united states
  government, 2017.

\bibitem{xforce}
IBM.
\newblock Ibm x-force exchange.
\newblock \url{https://exchange.xforce.ibmcloud.com/}.

\bibitem{d2canvas}
{Immunity Inc.}
\newblock Canvas.
\newblock \url{https://www.immunityinc.com/products/canvas/}, 2019.

\bibitem{WEIS-19-Exploits}
J.~Jacobs, S.~Romanosky, I.~Adjerid, and W.~Baker.
\newblock Improving vulnerability remediation through better exploit
  prediction.
\newblock In {\em The 2019 Workshop on the Economics of Information Security
  (WEIS)}, Jun 2019.

\bibitem{jacobs2019exploit}
J.~Jacobs, S.~Romanosky, B.~Edwards, I.~Adjerid, and M.~Roytman.
\newblock Exploit prediction scoring system (epss).
\newblock {\em Digital Threats: Research and Practice}, 2(3), July 2021.

\bibitem{landman2016empirical}
D.~Landman, A.~Serebrenik, E.~Bouwers, and J.~J. Vinju.
\newblock Empirical analysis of the relationship between cc and sloc in a large
  corpus of java methods and c functions.
\newblock {\em Journal of Software: Evolution and Process}, 28(7):589--618,
  2016.

\bibitem{le2014distributed}
Q.~Le and T.~Mikolov.
\newblock Distributed representations of sentences and documents.
\newblock In {\em International conference on machine learning}, pages
  1188--1196. PMLR, 2014.

\bibitem{li2017large}
F.~Li and V.~Paxson.
\newblock A large-scale empirical study of security patches.
\newblock In {\em Proceedings of the 2017 ACM SIGSAC Conference on Computer and
  Communications Security}, pages 2201--2215, 2017.

\bibitem{liu2015cloudy}
Y.~Liu, A.~Sarabi, J.~Zhang, P.~Naghizadeh, M.~Karir, M.~Bailey, and M.~Liu.
\newblock Cloudy with a chance of breach: Forecasting cyber security incidents.
\newblock In {\em 24th USENIX Security Symposium (USENIX Security 15)}, pages
  1009--1024, 2015.

\bibitem{lopez2016revisiting}
D.~Lopez-Paz and M.~Oquab.
\newblock Revisiting classifier two-sample tests.
\newblock {\em arXiv preprint arXiv:1610.06545}, 2016.

\bibitem{mann2007simple}
G.~S. Mann and A.~McCallum.
\newblock Simple, robust, scalable semi-supervised learning via expectation
  regularization.
\newblock In {\em Proceedings of the 24th international conference on Machine
  learning}, pages 593--600, 2007.

\bibitem{melis2018explaining}
M.~Melis, D.~Maiorca, B.~Biggio, G.~Giacinto, and F.~Roli.
\newblock Explaining black-box android malware detection.
\newblock In {\em 2018 26th European Signal Processing Conference (EUSIPCO)},
  pages 524--528. IEEE, 2018.

\bibitem{menon2016learning}
A.~K. Menon, B.~Van~Rooyen, and N.~Natarajan.
\newblock Learning from binary labels with instance-dependent corruption.
\newblock {\em arXiv preprint arXiv:1605.00751}, 2016.

\bibitem{perlparser}
{MetaCPAN}.
\newblock Compiler::parser.
\newblock \url{https://metacpan.org/pod/Compiler::Parser}, 2020.

\bibitem{contagio}
{Mila Parkour}.
\newblock Contagio dump.
\newblock
  \url{http://contagiodump.blogspot.com/2010/06/overview-of-exploit-packs-update.html},
  2019.

\bibitem{eternalblue}
{MITRE CVE}.
\newblock Cve-2017-0144.
\newblock \url{https://cve.mitre.org/cgi-bin/cvename.cgi?name=CVE-2017-0144},
  2019.

\bibitem{mu2018understanding}
D.~Mu, A.~Cuevas, L.~Yang, H.~Hu, X.~Xing, B.~Mao, and G.~Wang.
\newblock Understanding the reproducibility of crowd-reported security
  vulnerabilities.
\newblock In {\em 27th {USENIX} Security Symposium ({USENIX} Security 18)},
  pages 919--936, Baltimore, MD, Aug. 2018. {USENIX} Association.

\bibitem{Nappa15:VulnerabilityPatching}
A.~Nappa, R.~Johnson, L.~Bilge, J.~Caballero, and T.~Dumitra\cb{s}.
\newblock The attack of the clones: A study of the impact of shared code on
  vulnerability patching.
\newblock In {\em S\&P}, 2015.

\bibitem{cvss3guide}
A complete guide to the common vulnerability scoring system.
\newblock \url{https://www.first.org/cvss/v3.0/specification-document}.

\bibitem{nvd}
National vulnerability database.
\newblock \url{http://nvd.nist.gov/}.

\bibitem{otx}
Alienvault otx.
\newblock AlienVault, 30 March 2009.
\newblock \url{https://otx.alienvault.com/}.

\bibitem{patrini2017making}
G.~Patrini, A.~Rozza, A.~Krishna~Menon, R.~Nock, and L.~Qu.
\newblock Making deep neural networks robust to label noise: A loss correction
  approach.
\newblock In {\em Proceedings of the IEEE Conference on Computer Vision and
  Pattern Recognition}, pages 1944--1952, 2017.

\bibitem{pendlebury2019tesseract}
F.~Pendlebury, F.~Pierazzi, R.~Jordaney, J.~Kinder, and L.~Cavallaro.
\newblock $\{$TESSERACT$\}$: Eliminating experimental bias in malware
  classification across space and time.
\newblock In {\em 28th $\{$USENIX$\}$ Security Symposium ($\{$USENIX$\}$
  Security 19)}, pages 729--746, 2019.

\bibitem{qualys:wannacry}
Massive microsoft patch tuesday security update for march.
\newblock Qualys, 30 March 2017.
\newblock
  \url{https://blog.qualys.com/laws-of-vulnerabilities/2017/03/14/massive-security-update-from-microsoft-for-march}.

\bibitem{redmonmag:wannacry}
Microsoft resumes security updates with 'largest' patch tuesday release.
\newblock Redmont Mag, 30 March 2017.
\newblock
  \url{https://redmondmag.com/articles/2017/03/14/march-2017-security-updates.aspx}.

\bibitem{pythonparser}
{Python}.
\newblock ast.
\newblock \url{https://docs.python.org/2/library/ast.html}, 2020.

\bibitem{metasploit}
{Rapid7}.
\newblock The metasploit framework.
\newblock \url{https://www.metasploit.com/}, 2019.

\bibitem{RedHat:SeverityRating}
Severity ratings.
\newblock RedHat, 30 March 2009.
\newblock \url{https://access.redhat.com/security/updates/classification/}.

\bibitem{Reuters:ExploitabilityIndexCritique}
Microsoft correctly predicts reliable exploits just 27
\newblock Reuters, 30 March 2009.
\newblock
  \url{https://www.reuters.com/article/urnidgns852573c400693880002576630073ead6/microsoft-correctly-predicts-reliable-exploits-just-27-of-the-time-idUS186777206820091104}.

\bibitem{rms_risk}
First probabilistic cyber risk model launched by RMS.
\newblock
  \url{https://www.artemis.bm/news/first-probabilistic-cyber-risk-model-launched-by-rms/}.

\bibitem{rubyparser}
{Ruby}.
\newblock Ripper.
\newblock
  \url{https://ruby-doc.org/stdlib-2.5.1/libdoc/ripper/rdoc/Ripper.html}, 2020.

\bibitem{USENIX-Security-2015}
C.~Sabottke, O.~Suciu, and T.~Dumitra\cb{s}.
\newblock Vulnerability disclosure in the age of social media: Exploiting
  {Twitter} for predicting real-world exploits.
\newblock In {\em USENIX Security Symposium}, Washington, DC, Aug 2015.

\bibitem{riskbasedsecurity:CVSS1}
R.~B. Security.
\newblock Cvssv3: New system, new problems (file-based attacks).
\newblock
  \url{https://www.riskbasedsecurity.com/2017/01/16/cvssv3-new-system-new-problems-file-based-attacks/}.

\bibitem{shahzad:12}
M.~Shahzad, M.~Z. Shafiq, and A.~X. Liu.
\newblock A large scale exploratory analysis of software vulnerability life
  cycles.
\newblock In {\em Proceedings of the 2012 International Conference on Software
  Engineering}, 2012.

\bibitem{skybox}
{SkyBox}.
\newblock Vulnerability center.
\newblock \url{https://www.vulnerabilitycenter.com/#home}, 2019.

\bibitem{DBLP:conf/uss/SoskaC14}
K.~Soska and N.~Christin.
\newblock Automatically detecting vulnerable websites before they turn
  malicious.
\newblock In {\em Proceedings of the 23rd {USENIX} Security Symposium, San
  Diego, CA, USA, August 20-22, 2014.}, pages 625--640, 2014.

\bibitem{sym_threatexplorer}
{Symantec Corporation}.
\newblock Symantec threat explorer.
\newblock \url{https://www.symantec.com/security-center/a-z}, 2019.

\bibitem{szekeres2013sok}
L.~Szekeres, M.~Payer, T.~Wei, and D.~Song.
\newblock Sok: Eternal war in memory.
\newblock In {\em 2013 IEEE Symposium on Security and Privacy}, pages 48--62.
  IEEE, 2013.

\bibitem{talosintelligence_wannacry}
Player 3 has entered the game: say hello to 'wannacry'.
\newblock \url{https://blog.talosintelligence.com/2017/05/wannacry.html}.

\bibitem{Tavabi2018DarkEmbedEP}
N.~Tavabi, P.~Goyal, M.~Almukaynizi, P.~Shakarian, and K.~Lerman.
\newblock Darkembed: Exploit prediction with neural language models.
\newblock In {\em AAAI}, 2018.

\bibitem{tenable}
{Tenable}.
\newblock Tenable research advisories.
\newblock \url{https://www.tenable.com/security/research}, 2019.

\bibitem{Tenable:Nessus}
{Tenable Network Security}.
\newblock {Nessus} vulnerability scanner.
\newblock \url{http://www.tenable.com/products/nessus}.

\bibitem{Tenable:RiskRating}
Severity vs. vpr.
\newblock Tenable, 30 March 2019.
\newblock \url{https://docs.tenable.com/tenablesc/Content/RiskMetrics.htm}.

\bibitem{Twitter:api}
{Twitter}.
\newblock Filtered stream.
\newblock
  \url{https://developer.twitter.com/en/docs/twitter-api/tweets/filtered-stream/introduction}.

\bibitem{virustotal}
Virustotal.
\newblock Virus total.
\newblock \url{www.virustotal.com}.

\bibitem{vulners}
Vulners.
\newblock Vulners vulnerability database.
\newblock \url{https://vulners.com/}.

\bibitem{IBM:RiskRating}
Y.~Watanabe.
\newblock Assessing security risk of your containers with vulnerability
  advisor.
\newblock IBM, 30 March 2019.
\newblock
  \url{https://medium.com/ibm-cloud/assessing-security-risk-of-your-containers-with-vulnerability-advisor-f6e45fff82ef}.

\bibitem{wu2018fuze}
W.~Wu, Y.~Chen, J.~Xu, X.~Xing, X.~Gong, and W.~Zou.
\newblock $\{$FUZE$\}$: Towards facilitating exploit generation for kernel
  use-after-free vulnerabilities.
\newblock In {\em 27th USENIX Security Symposium (USENIX Security 18)}, pages
  781--797, 2018.

\bibitem{xiao2018patching}
C.~Xiao, A.~Sarabi, Y.~Liu, B.~Li, M.~Liu, and T.~Dumitras.
\newblock From patching delays to infection symptoms: using risk profiles for
  an early discovery of vulnerabilities exploited in the wild.
\newblock In {\em 27th USENIX Security Symposium (USENIX Security 18)}, pages
  903--918, 2018.

\bibitem{xiao2015learning}
T.~Xiao, T.~Xia, Y.~Yang, C.~Huang, and X.~Wang.
\newblock Learning from massive noisy labeled data for image classification.
\newblock In {\em Proceedings of the IEEE conference on computer vision and
  pattern recognition}, pages 2691--2699, 2015.

\bibitem{yamaguchi2014modeling}
F.~Yamaguchi, N.~Golde, D.~Arp, and K.~Rieck.
\newblock Modeling and discovering vulnerabilities with code property graphs.
\newblock In {\em 2014 IEEE Symposium on Security and Privacy}, pages 590--604.
  IEEE, 2014.

\bibitem{zdnet-win10disclosureexploit}
{ZDNet}.
\newblock Recent windows alpc zero-day has been exploited in the wild for
  almost a week.
\newblock
  \url{https://www.zdnet.com/article/recent-windows-alpc-zero-day-has-been-exploited-in-the-wild-for-almost-a-week/},
  2018.

\end{thebibliography}
}

\appendix
\section{Appendix}

\begin{figure}[th]
    \begin{subfigure}{.24\textwidth}
        \centering
        \includegraphics[height=1.8in]{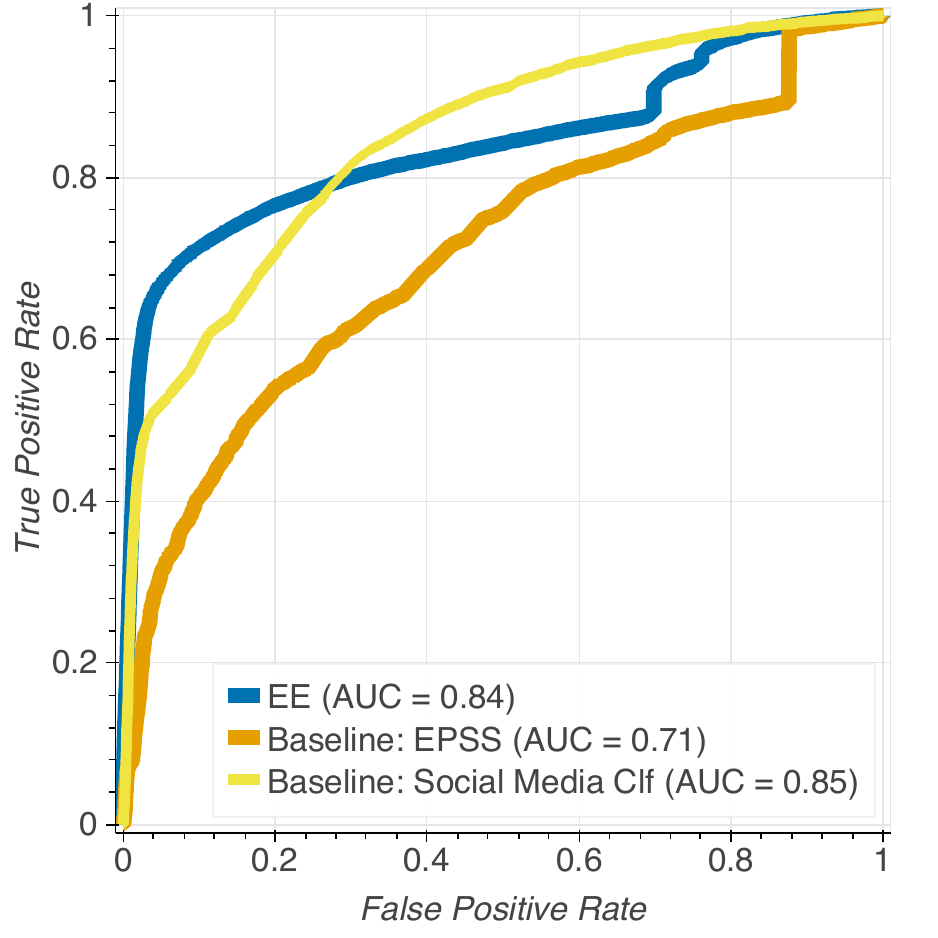}
        \vspace{-0.2in}
        \caption{}
        \label{fig:perf_a_appendix}
    \end{subfigure}%
    \begin{subfigure}{.24\textwidth}
        \centering
        \includegraphics[height=1.8in]{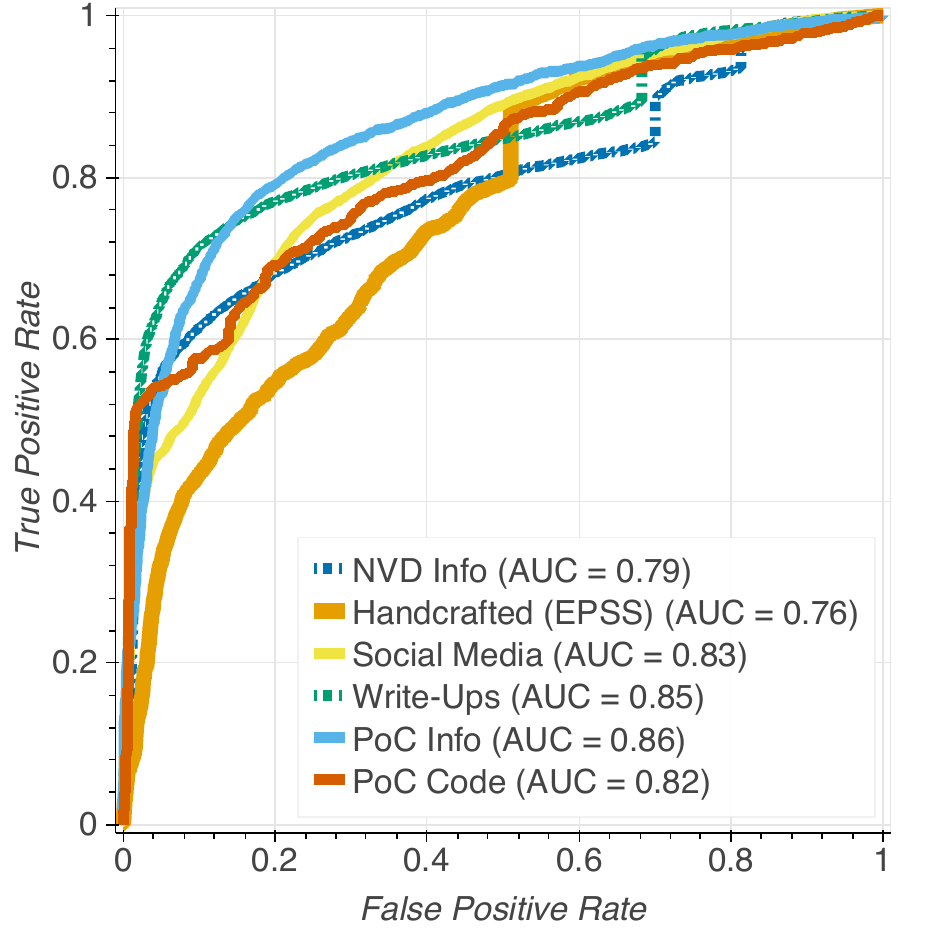}
        \vspace{-0.2in}
        \caption{}
        \label{fig:perf_b_appendix}
    \end{subfigure}%
    \caption{ROC curves for the corresponding precision-recall curves in Figure~\ref{fig:prediction_performance1}.}
    \label{fig:perf_ab_appendix}
\end{figure}

\begin{figure}[th]
\centering
\includegraphics[height=1.8in]{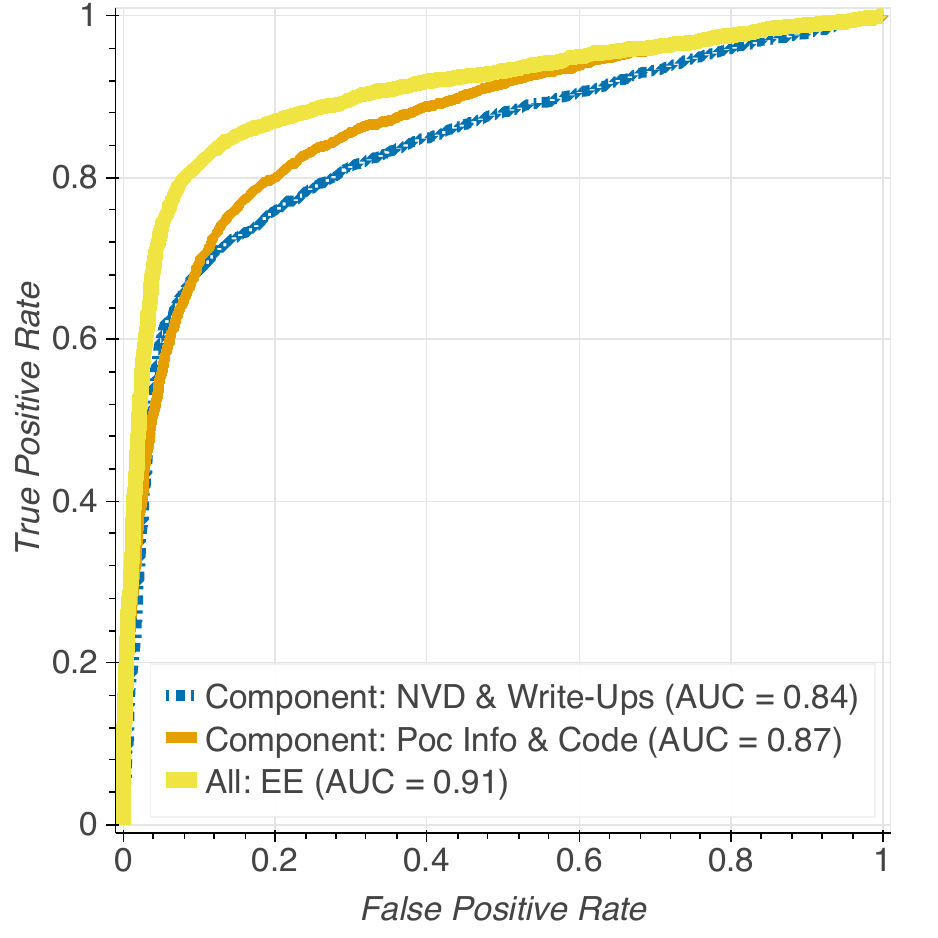}
\caption{ROC curve for the corresponding precision-recall curves in Figure~\ref{fig:prediction_performance2a}.}
\label{fig:perf_c_appendix}
\end{figure}

\begin{figure}
\vspace{-0.5in}
    \begin{subfigure}{.24\textwidth}
        \centering
        \includegraphics[height=1.8in]{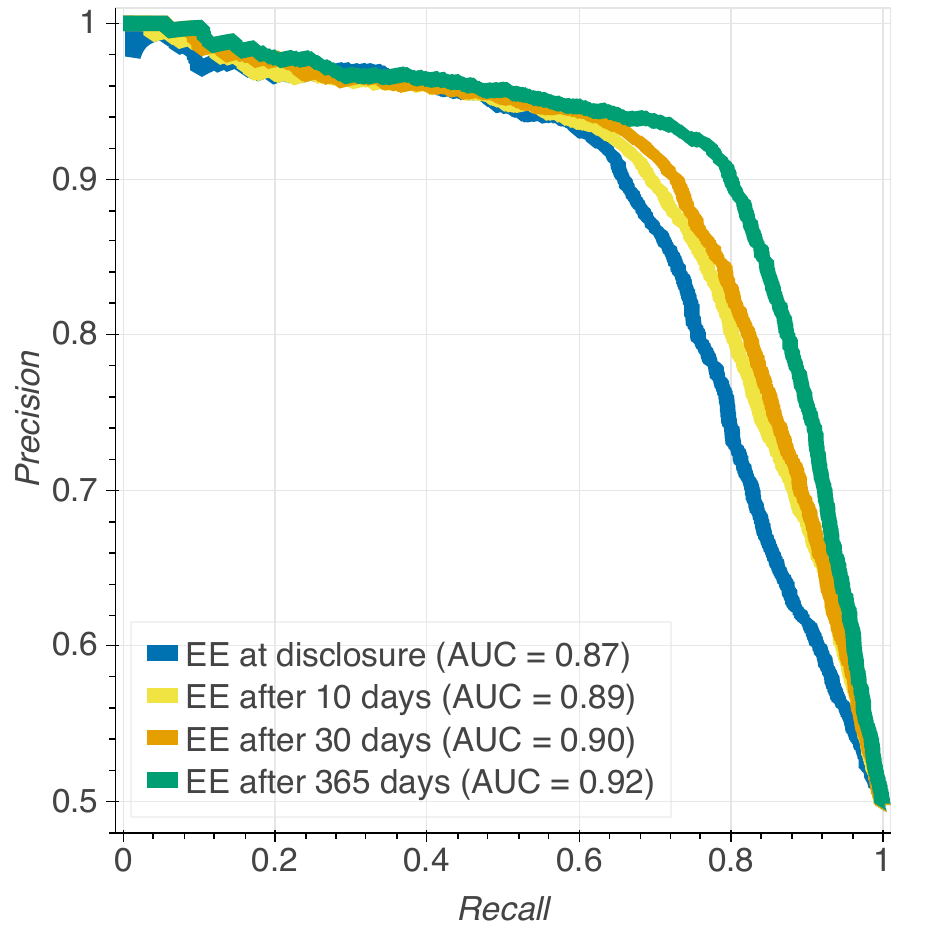}
        \vspace{-0.2in}
        \caption{}
        \label{fig:{perf_d}}
    \end{subfigure}%
    \begin{subfigure}{.24\textwidth}
        \centering
        \includegraphics[height=1.8in]{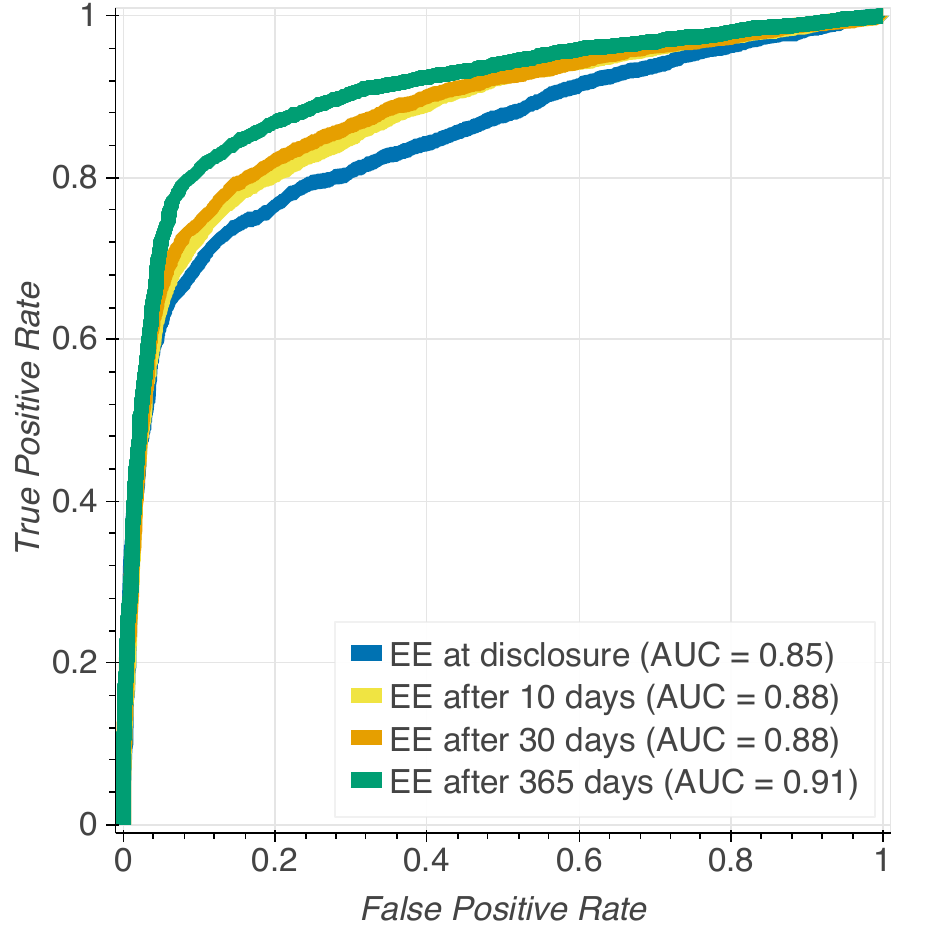}
        \vspace{-0.2in}
        \caption{}
        \label{fig:perf_d_appendix}
    \end{subfigure}%
    \caption{Performance of \texttt{EE} evaluated at different points in time.}
    \label{fig:prediction_performance_over_time}
\end{figure}

\subsection{Evaluation}
\label{appendix:a1}

\topic{Additional ROC Curves} 
Figures~\ref{fig:perf_ab_appendix} and ~\ref{fig:perf_c_appendix} highlight the trade-offs between true positives and false positives in classification.

\topic{\texttt{EE} performance improves over time}
To observe how our classifier performs over time, in Figure~\ref{fig:prediction_performance_over_time} we plot the performance when \texttt{EE} is computed at disclosure, then 10, 30 and 365 days later.
We observe that the highest performance boost happens within the first 10 days after disclosure, where the AUC increases from 0.87 to 0.89. 
Overall, we observe that the performance gains are not as large later on: the AUC at 30 days being within 0.02 points of that at 365 days.
This suggests that the artifacts published within the first days after disclosure have the highest predictive utility, and that the predictions made by \texttt{EE} close to disclosure can be trusted to deliver a high performance.

\section{Artifact Appendix}
\label{artifact_appendix:0}

\subsection{Abstract}

We developed a Web platform and an API client that allow users to retrieve the Expected Exploitability (\texttt{EE}) scores predicted by our system.
The system gets updated daily with the newest scores.

We implemented an API client in python, distributed via Jupyter notebooks in a Docker container, which allows users to interact with the API and download the \texttt{EE} scores to reproduce the main result from the paper, in \textbf{Figure~\ref{fig:prediction_performance1a}} and \textbf{Figure~\ref{fig:perf_a_appendix}}, or explore the performance of the latest model and compare it to the performance of the models from the paper.

\subsection{Artifact check-list (meta-information)}

{\small
\begin{itemize}
  \item {\bf Program: } Docker (tested on v20.10.8) 
  \item {\bf Run-time environment: } UNIX-like system
  \item {\bf Metrics: } Precision, Recall, Precision-Recall AUC, TPR, FPR, AUC.
  \item {\bf Output: } Plots from the paper, reproducing and expanding Figure~\ref{fig:prediction_performance1a} and Figure~\ref{fig:perf_a_appendix}.
  \item {\bf Experiments: } Running Jupyter notebooks.
  \item {\bf How much disk space required (approximately)?: } 4GB
  \item {\bf How much time is needed to prepare workflow (approximately)?: } 15 min
  \item {\bf How much time is needed to complete experiments (approximately)?: } 30 min
  \item {\bf Publicly available?: } The website and client code are publicly available. The API requires an API token which we provide upon request. 
\end{itemize}

\subsection{Description}

\subsubsection{Web Platform}

The Web platform exposes the scores of the most recent model, and offers two tools for practitioners to integrate \texttt{EE} in vulnerability or risk management workflows.

The \textit{Vulnerability Explorer} tool allows users to search and investigate basic characteristics of any vulnerability on our platform, the historical scores for that vulnerability as well as a sample of the artifacts used in computing its \texttt{EE}.
One use-case for this tool is the investigation of critical vulnerabilities, as discussed in Section~\ref{sec:casestudy} - \texttt{EE} for critical vulnerabilities.

The \textit{Score Comparison} tool allows users to compare the scores across subsets of vulnerabilities of interest.
Vulnerabilities can be filtered based on the publication date, type, targeted product or affected vendor.
The results are displayed in a tabular form, where users can rank vulnerabilities according to various criteria of interest (e.g., the latest or maximum \texttt{EE} score, the score percentile among selected vulnerabilities, whether an exploit was observed etc.).
One use-case for the tool is the discovery of critical vulnerabilities that need to be prioritized soon or for which exploitation is imminent, as discussed in Section~\ref{sec:casestudy} - \texttt{EE} for emergency response.

\subsubsection{API Client} 

The API allows clients to download historical scores for a given vulnerability (using the \texttt{/scores/cveid} endpoint), or all the prediction scores for a particular model on a particular date (using the \texttt{/scores/daily} endpoint).
The API documentation describes the endpoints and the parameters required for each call, and provides example code for clients to interact with the API.

\subsection{How to access}

The Web platform is available at \href{https://exploitability.app/}{https://exploitability.app/}.
The API and the client code are available at \href{https://api.exploitability.app/}{https://api.exploitability.app/}.

\subsection{Installation}

The practitioner tools are available on the Web platform. 
To use the API, clone the code repository, point a terminal to that folder and run \texttt{bash docker/run.sh}. This will create the Docker container and spawn a Jupyter server.
Use the URL displayed in the console to open a browser session within that container.

\subsection{Evaluation and expected results}

To reproduce the results from the paper, open and run the following notebook:
 \texttt{reproducibility\_plot\_performance.ipynb}
In an API key is provided, this will download the required scores used in the paper, cache them in various files in  \texttt{scores\_reproducibility\_download/}, and use these files to compute the performance of \texttt{EE} and baselines.
The output consists of 2 figures, which correspond to Figure~\ref{fig:perf_a_appendix} and Figure~\ref{fig:prediction_performance1a} in our paper.

To evaluate the latest model, open and run the following notebook:
\texttt{latest\_plot\_performance.ipynb}
In an API key is provided, the notebook will download all the scores produced by our latest model on \texttt{2021-10-10}, cache them into a file in \texttt{scores\_latest\_download/}, and use this file to compute the performance of \texttt{EE}.
The output consists of 2 figures which are comparable to Figure~\ref{fig:perf_a_appendix} and Figure~\ref{fig:prediction_performance1a} in our paper.

As of December 2021, we observe that the performance of our latest model predicting \texttt{EE}, computed before \texttt{2021-10-10}, is very close to the performance reported in the paper (0.69 PR AUC / 0.96 AUC for latest model vs 0.73 PR AUC / 0.84 AUC in the paper), demonstrating that our online predictor is functional and in line with the claims from the paper.
The performance of the latest model on other dates can be computed by changing the \texttt{SCORES\_DATE} variable and re-running the notebook.

\subsection{Experiment customization}

When running \texttt{latest\_plot\_performance.ipynb}, users can customize the \texttt{SCORES\_DATE} variable in the notebook to observe the performance of our model on different dates.
\end{document}